\documentclass[10pt, article, amsmath, onecolumn, showpacs, superscriptaddress]{revtex4-1}
\usepackage{graphicx}
\usepackage[latin1]{inputenc}
\usepackage{hyperref}
\usepackage{amsfonts}
\usepackage[normalem]{ulem}
\usepackage{bm}
\usepackage{color}
\definecolor{dgreen}{rgb}{0,0.7,0}
\def\nn{\nonumber}

\def\la{\langle}
\def\ra{\rangle}
\def\ql{ {Q_{ l}} }
\def\qr{ {Q_{ r}} }
\usepackage{tikz}
\usetikzlibrary{shapes,shapes.multipart}

\usepackage[mathscr]{euscript}

\newcommand{\f}{\frac}
\newcommand\bea{\begin{eqnarray}}
\newcommand\eea{\end{eqnarray}}
\newcommand\p{\partial}
\newcommand\ie{{\emph{i.e.}}}

\newcommand{\ket}[1]{ |\;#1\;\rangle}
\newcommand{\bra}[1]{ \langle\;#1\;|}
\newcommand{\iprod}[2]{ \langle#1|#2\rangle}

\newcommand{\mc}[1]{\mathcal{#1}}
\newcommand{\mb}[1]{\mathbb{#1}}

\newcommand{\dx}[1]{\hat{#1}}
\newcommand{\blang}{\big \langle}
\newcommand{\brang}{\big \rangle}

\newcommand{\flapfinitem}{\mb{L}^p}

\def\bea{\begin{eqnarray}}
\def\eea{\end{eqnarray}}

\newcommand{\eqa}[1]{\begin{align}#1\end{align}}

\newcommand{\iu}{{i\mkern1mu}}

\newcommand{\RN}[1]{
	\textup{\uppercase\expandafter{\romannumeral#1}}%
}

\newcommand{\dd}{\mathrm{d}}

\newcommand{\cs}{c_\mathrm{s}}

\newcommand{\Tp}{T_+}
\newcommand{\Tm}{T_-}

\newcommand{\Tb}{\overline{T}}

\newcommand{\phip}{\phi_+}
\newcommand{\phim}{\phi_-}
\newcommand{\phipm}{\phi_\pm}
\newcommand{\phiz}{\phi_0}

\newcommand{\etapm}{\eta_\pm}

\begin{document}
\title{Anomalous heat transport in one dimensional systems: a description using non-local fractional-type diffusion equation}
\author{Abhishek Dhar}
\affiliation{International Centre for Theoretical Sciences, Tata Institute of Fundamental Research, Bengaluru 560089, India}
\author{Anupam Kundu} 
\affiliation{International Centre for Theoretical Sciences, Tata Institute of Fundamental Research, Bengaluru 560089, India}
\author{Aritra Kundu}
\affiliation{Raman Research Institute, Bengaluru 560012, India}
\date{\today}

\begin{abstract}
It has been observed in many numerical simulations, experiments and from various theoretical treatments that heat transport in  one-dimensional systems of interacting particles  cannot be described by the phenomenological  Fourier's law. 
The picture that has emerged from studies over the last  few years is that Fourier's law gets replaced by a spatially non-local  linear  equation wherein the current at a point gets contributions from the temperature gradients in other parts of the system. Correspondingly the usual heat diffusion equation gets replaced by a non-local fractional-type diffusion equation.  
In this review, we describe the various theoretical approaches which lead to this framework  and also discuss recent progress on this problem.
\end{abstract}

\maketitle

\section{Introduction}
\noindent
Transport of heat through materials is a paradigmatic example of non-equilibrium phenomena \cite{Bonetto00,Lepri03,Dhar08}. When an extended system is attached to two reservoirs of different temperatures at its two ends, an energy current  flows through the body from hot region to cold region. At the macroscopic level this phenomena is described by the phenomenological Fourier's law. Considering transport in one dimensional systems, Fourier's law states that the local heat current density $j(x,t)$ inside a system at point $x$ at time $t$ is proportional to the gradient of the local temperature  $T(x,t)$: 
\begin{align}
j = - \kappa \frac{\partial T(x,t)}{\partial x}~~~\label{FL}
\end{align}
where  $\kappa$ is referred to as the thermal conductivity of the material. This law implies diffusive transfer of energy. To see this we  note that  the local energy density $e(x,t)$ in a one dimensional system  satisfies the continuity equation ${\partial e(x,t)}/{\partial t} =- {\partial j(x,t)}/{\partial x}$. Inserting Eq.~\eqref{FL} in this continuity equation, and using  the relation between the local energy density and the local temperature $c_v=\partial e/\partial T$ (where $c_v$ represents the specific heat per unit volume), 
one finds the heat diffusion equation
\begin{align}
\frac{\partial T(x,t)}{\partial t} = \frac{\kappa }{c_v}  \frac{\partial^2 T(x,t)}{\partial x^2}, \label{diff-T-HD} 
\end{align}
where we assume (for simplicity) no variation of $\kappa$ with temperature. In usual three dimensional systems, the heat diffusion equation takes the form $\p_t T({\bf x},t)=({\kappa }/{c_v}) \nabla^2 T({\bf x},t)$ and describes the evolution of the temperature field in bulk systems. The phenomenological macroscopic description provided by the equations in \eqref{FL} and \eqref{diff-T-HD} has been used extensively to describe heat transfer phenomena in a wide class of physical systems.

A natural question is to ask if it is possible to derive or establish Fourier's  phenomenological law theoretically, starting from a complete microscopic description.  The issue of deriving Fourier's law has been a long standing question and a very active  field of research  \cite{Bonetto00}.  Several theoretical as well as large scale numerical studies have been performed on different  mathematical model systems to understand the necessary and sufficient conditions needed in the microscopic description to validate Fourier's law at the macroscopic level \cite{Lepri03, Dhar08, Lepri16}.  
Surprisingly, these studies suggest that  Fourier's law is probably not valid in many one-dimensional systems and one finds that the thermal conductivity $\kappa$  diverges  with  system size $N$ as $\kappa \sim N^\alpha$ where $0 < \alpha < 1$ \cite{Lepri97,Dhar01,Grassberger02,Lepri03,Cipriani05,Basile06,Dhar08,Lepri16, Mai07,Dhar08a,Chen14}. This is referred to as {\bf anomalous heat transport} (AHT). For $\alpha=0$, the transport  is classified as being diffusive while  $\alpha=1$ is referred to as ballistic transport \cite{Lepri03, Dhar08}. 
Recent developments in technology has made it possible to verify some of these theoretical predictions experimentally  as well as numerically in real physical systems such as nano-structures, polymers, semiconductor films etc. \cite{ Chang08, Xu14, Lee17, Majee16,Pereira13, Nika12, Xu16,Meier14}, and these have provided further motivation and new directions of study.  

 Two approaches have mainly been used to look for signatures of anomalous heat transport (AHT): (i) the {\bf open system} set-up in which a system is connected to heat reservoirs at different temperatures $T_L$ and $T_R$ at the two ends and  (ii) the {\bf closed system} set-up in which the isolated system is prepared in thermal equilibrium at temperature $T$ and evolves according to  Hamiltonian dynamics (or sometimes stochastic dynamics with same conservation laws). In the open system set-up, one usually  considers the  non-equilibrium steady state (NESS) and measures directly the steady state heat current $j$  and the temperature profile $T(x)$ in a finite system of $N$ particles. For small $\Delta T=T_L-T_R$, one finds the system size scaling $j \sim N^{\alpha-1}$ (implying $\kappa \sim N^\alpha$) and a \emph{non-linear} temperature profile. These are in contrast with Fourier's law which would predict $j \sim N^{-1}$ and a linear temperature profile. In the closed system set-up the idea is to look at the spreading of a heat pulse in a system in equilibrium. From linear response theory we expect that this would evolve in the same way as dynamical correlations of energy fluctuations in equilibrium. 
Studies on  spreading of pulses and energy correlations in systems with AHT show that the process is super-diffusive, with scaling functions described by  L\'{e}vy distributions \cite{Cipriani05,Zhao06,Mendl13}. This contrasts systems described by Fourier's law where we expect diffusion and Gaussian propagators. Note that 
we expect in fact that the thermal conductivity $\kappa$ obtained in non-equilibrium measurements should be related to equilibrium energy current auto-correlation functions via the Green-Kubo formula  \cite{Green54, Kubo57, Dhar08}. This leads to the  understanding of AHT as arising from the fact that the non-integrable long time tails in the auto-correlation function of the total current lead to the divergence of the thermal conductivity. 

The natural question that arises for understanding systems with AHT is to find the replacements of Fourier's law in Eq.~\eqref{FL} and the heat diffusion equation in  Eq.~\eqref{diff-T-HD}. The picture that has emerged from studies over the last  few years \cite{Jara09, Lepri09, Lepri16, Lepri10, Delfini10, Jara15, Bernardin12, Bernardin16, Mellet11, Basile09, Kundu18, Miron19, Cividini17, Kundu19} is that Fourier's law gets replaced by a spatially non-local  but linear equation wherein the current at a point gets contributions from temperature gradients in other parts of the system. This has the form
\begin{align}
j(x,t)=-\int dx'~K(x,x')~\frac{\p T(x',t)}{\p x'},~~\label{NLFL}
\end{align} 
where now the thermal conductivity is replaced by the non-local kernel $K(x,x')$.  This then leads to a corresponding non-local fractional-type equation for the time evolution of $T(x,t)$. An important difference from the heat diffusion equation is that the fractional-type equation takes different forms in the closed system set-up (infinite  domain) and the open system set-up (finite domain). 
 In the infinite domain the evolution of a localized temperature pulse is described by a  fractional-type diffusion equation  
\begin{align} \label{fraceq}
\partial_t T(x,t) = -\bar{\kappa}(-\Delta)^{\nu/2} T(x,t),
\end{align}
where  the fractional operator should be interpreted by its action on plane wave basis states: $(-\Delta)^{\nu/2} e^{i k x}=|k|^{\nu}~e^{ik x}$, with $1 <\nu <2$.  
However it should be noted that the correpsonding  L\'{e}vy-stable distribution is valid only over the scale $x \lesssim t^{1/\nu}$. As we will see, the evolution of a heat pulse is restricted to a domain $|x|< ct$, determined by the sound speed $c$. For the open system, the precise form of the fractional equation is  dependent on  the details of boundary conditions. In this review we discuss these developments as well as open questions.

The plan of the review is as follows. In Sec.~(\ref{sec:signatures}) we discuss the various signatures of AHT in the closed and open set-ups. In Sec.~(\ref{sec:theory}) we discuss two theoretical approaches that have been used to understand various aspects of anomalous transport.  One of these is a phenomenological approach based on the idea that the heat carriers perform  L\'{e}vy walks instead of random walk. The second approach is a microscopic one, though still phenomenological, and is based on fluctuating hydrodynamics and applicable to Hamiltonian systems. For a class of stochastic models, it has been possible to provide a complete microscopic derivation of the fractional heat equation in the context of both the closed and open system set-ups. These results are described in Sec.~(\ref{sec:stochastic}). In the last part of this section we address the difficult issue of treating arbitrary boundary conditions and discuss a heuristic formulation that uses linear response ideas and fluctuating hydrodynamics to arrive at a general form of the  kernel $K(x,x')$ in Eq.~(\ref{NLFL}). Finally we conclude in Sec.~(\ref{sec:summary}) with a summary of the results presented and some of the outstanding open questions.

\begin{figure}[t]
\includegraphics[width=0.75\linewidth]{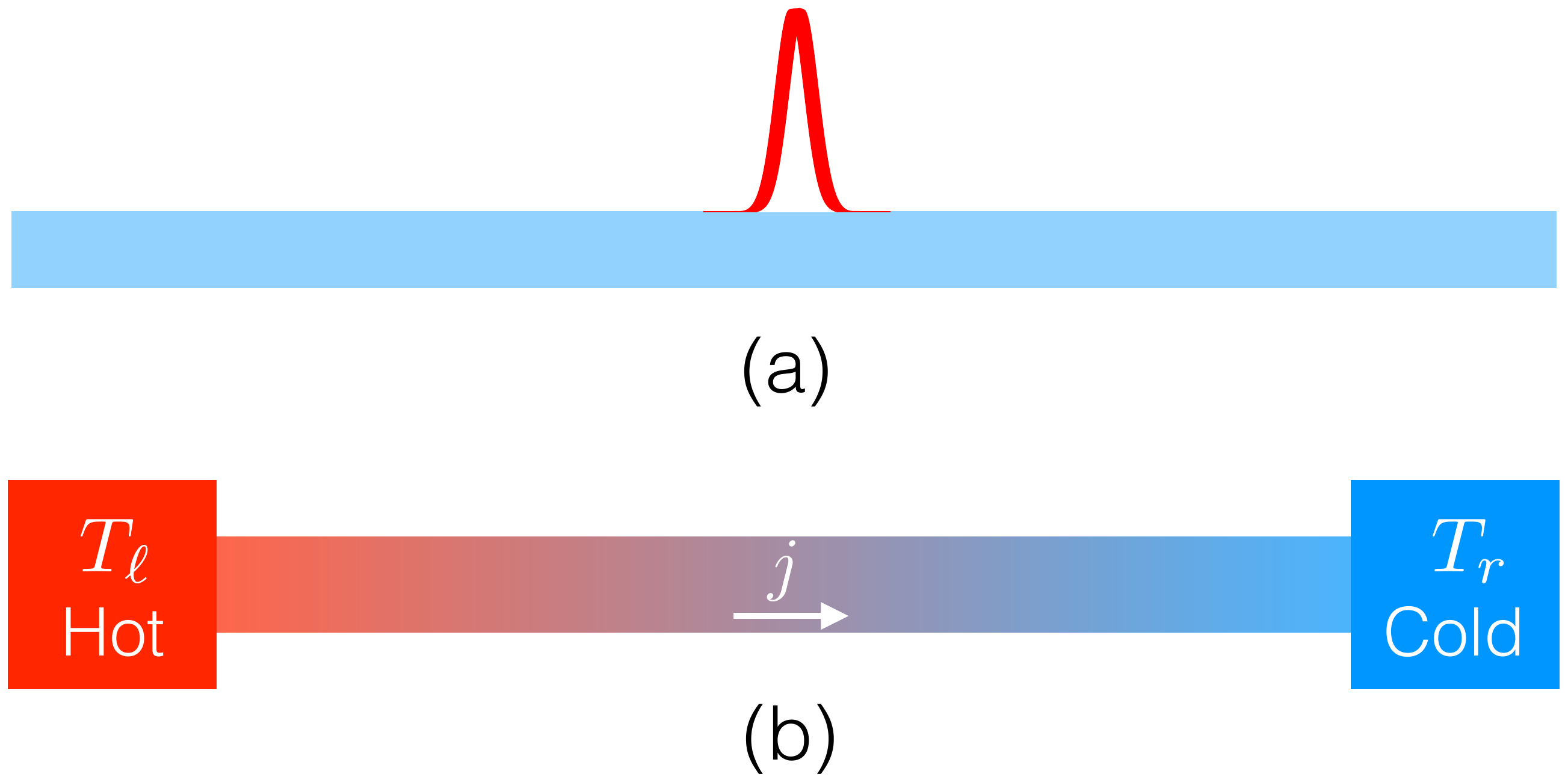}
\caption{ Schematic illustration of the (a) closed system set-up and the (b) open system set-up, commonly used to study heat transport. In (a), a localized heat pulse is  introduced at some point in a system in thermal equilibrium and its subsequent time-evolution is observed. In (b), the system is attached to two heat reservoirs at different temperatures and the NESS properties such as current and temperature profile are studied.} 
\label{mod}
\end{figure}

\section{Signatures of anomalous heat transport}
\label{sec:signatures}
\noindent
In the theoretical study of anomalous energy transport in one dimension, one usually considers simple yet non-trivial model systems of interacting particles. 
Let us consider $N$ particles of unit masses, with positions and momenta given, respectively, by $q_\ell$ and $p_\ell$, for $\ell=1,2,...,N$. One often starts with the following microscopic Hamiltonian:
\begin{equation}
H = \sum_{\ell=1}^{N}  \frac{p_\ell^2}{2} + \sum_{\ell=0}^{N} V(q_{\ell+1}-q_\ell), 
\label{hamiltonian}
\end{equation}
where $V(r)$ is a nearest neighbor interaction potential, and the extra variables $q_0$ and $q_{N+1}$ are introduced to incorporate different boundary conditions (BC). For example, fixed  BC corresponds to $q_0=0, q_{N+1}=0$ while free BC corresponds to setting $q_0=q_1,~q_{N+1}=q_N$.  The particles in the bulk of the system satisfy Hamiltonian equations of motion
\eqa{\label{eqmot}
\dot{q}_\ell=\p_{p_\ell}H,~\dot{p}_\ell=-\p_{q_\ell}H~,~~\ell=1,2,....N. 
}
One of the well-studied choices for the potential is to take $V(r)=k_2 r^2/2+k_3r^3/3+k_4 r^4/4$ which leads to the {\bf Fermi-Pasta-Ulam-Tsingou} (FPUT) model. Another popular choice is the {\bf alternate mass hard particle gas} which is not in the standard form of Eq.~\eqref{hamiltonian}. In this model one considers a chain of point particles with masses which alternate between two fixed values, say $m_1, m_2$, and which collide via elastic collisions conserving energy and momentum. 
For generic interaction potentials $V(r)$ it is expected that the system has three conserved quantities, namely volume of the system (alternatively the total number of particles), total momentum and total energy. Corresponding to each conserved quantity one can write a local continuity equation. For instance, the local energy defined on bulk points as 
\eqa{
e(\ell,t)=\frac{p_\ell^2}{2}+\frac{1}{2}[V(q_{\ell+1}-q_\ell)+V(q_\ell-q_{\ell-1})], 
}
satisfies a continuity equation
\eqa{\label{eq:jmic}
\p_t e(\ell,t)= j(\ell,t)-j({\ell+1},t),~~ {\rm where}~~j(\ell,t)=-\frac{1}{2}(p_{\ell-1}+p_\ell) V'(q_{\ell}-q_{\ell-1})~.
}
This equation  gives a microscopic definition of the energy current. 
 For quadratic $V(r)$, \emph{i.e} harmonic chains, there are a macroscopic number of conserved quantities and transport becomes ballistic. In this case a number of studies have considered augmenting the Hamiltonian dynamics with  a stochastic component such that the system again has only three conserved quantities \cite{Jara15, Basile06, Bernardin12, Bernardin16}. In this case one again recovers the typical features of anomalous transport and several exact results are possible. In this review we will discuss results for both Hamiltonian and stochastic systems. 
 
There are two possible approaches for studying transport properties of a system \cite{Dhar08,Lepri16}. A schematic of the two set-ups is shown in Fig.~(\ref{mod}):
\begin{itemize}
\item[{\bf A.}] Closed system set-up --- in this case, an isolated system is  prepared in thermal equilibrium  at  some temperature $T$ described by the canonical distribution
\eqa{
P({\bf q},{\bf p})=\frac{e^{-H({\bf q},{\bf p})/T}}{Z}~,
}
where $Z=\int d{\bf q}d{\bf p} e^{-H/T}$ is the partition function. For any initial condition chosen from this distribution the system evolves according to the pure Hamiltonian dynamics (or the conservative stochastic dynamics). Transport properties are usually probed by  studying the form of spatio-temporal correlation functions of the conserved quantities  (volume, momentum, energy)  or the decay with time of the energy current auto-correlation function. Another approach that has been used is to study the spreading of an initially  localized perturbation in the equilibrated system [see Fig.~(\ref{mod}a)].
In the closed system set-up one  takes the system to be infinite or, in numerical studies, $N$ to be sufficiently large such that the  correlations are not affected by the boundaries at the maximum observation times. \\
\item[{\bf B.}] Open system set-up ---  in this case, one considers finite systems attached at the two boundaries to  heat reservoirs  at different temperatures [see Fig.~(\ref{mod}b)].  The heat reservoirs are modeled by adding extra force  terms 
to the usual Hamiltonian equations of motion of the boundary particles. One of the standard choices is to consider Langevin type baths, wherein the additional forces consist of a dissipative term and a white noise term, which are related via a fluctuation-dissipation relation. The system is ``open'' in the sense that 
energy can flow  in and out of the system, though we note that locally in the bulk we still have energy conservation.  
When the temperatures of the heat reservoirs are different, the system eventually reaches a NESS in which a heat current flows across the system. The main focus of this approach has been to search for anomalous features in the NESS by looking at observables such as  the heat current $j=\la j(x,t) \ra^{neq}_{open}$ and temperature profile obtained from  $T(x)= \la p_x^2 \ra^{neq}_{open}$  (the averages are computed in the NESS). There have also been attempts to  understand the relaxation to NESS and look at correlations and large deviation properties of the NESS.
\end{itemize}

In the following sub-sections, we describe various signatures of AHT observed in both the set-ups.

\subsection{Signatures in the closed system set-up}
\label{sec:sign-closed}
\begin{figure}
\includegraphics[width=0.55\linewidth,angle=-90]{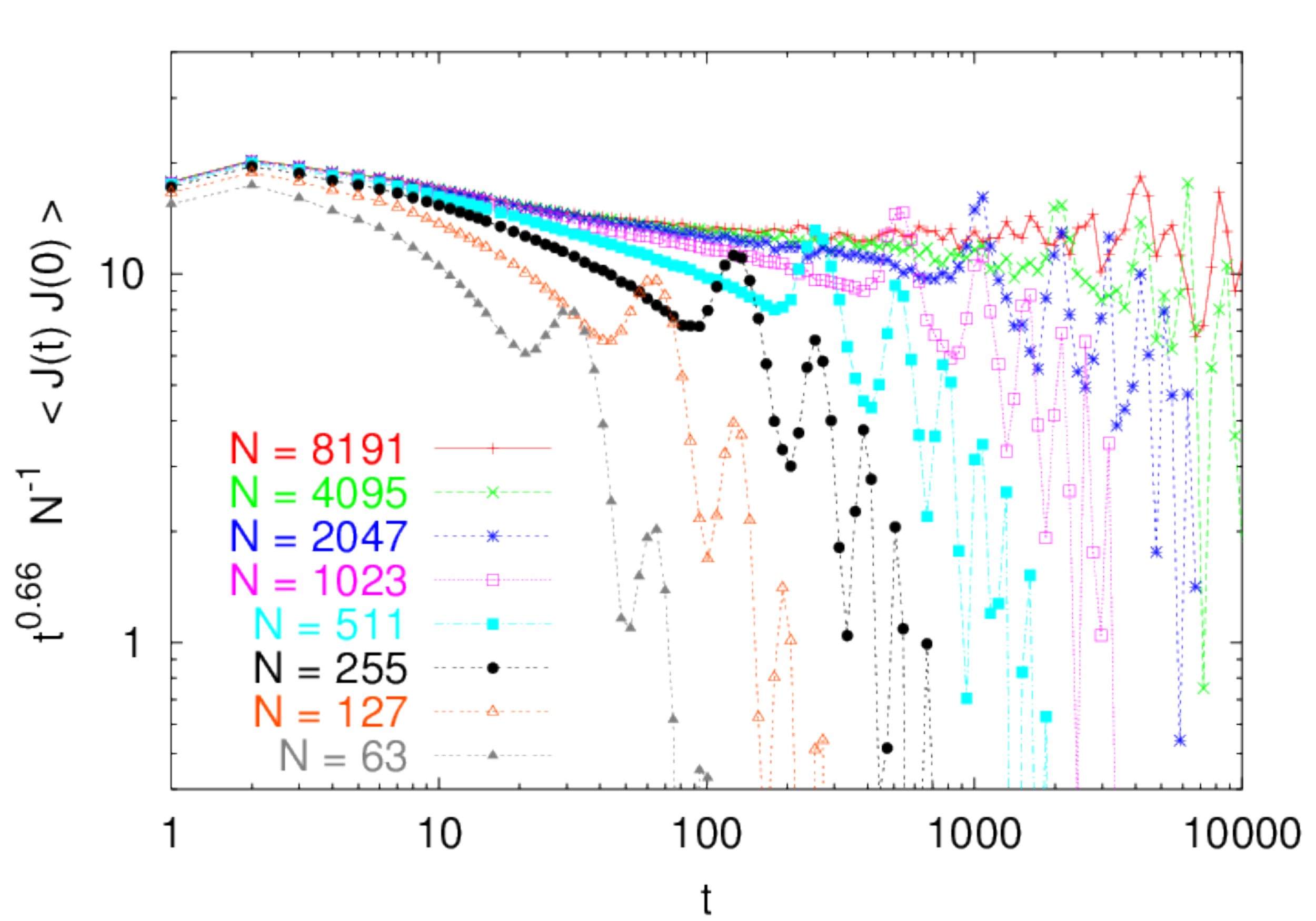}
\caption{Total scaled heat current auto-correlation, $t^{0.66} N^{-1} \la J(t)J(0)\ra$, in the alternate mass hard particle gas for mass ratio $2.2$ and $T=2.0$.  (Adapted from Grassberger et.al. with permission from  \cite{Grassberger02} Copyright (2002) by American Physical Society)}
\label{fig:grass}
\end{figure} 
\begin{figure}
\includegraphics[width=0.75\linewidth]{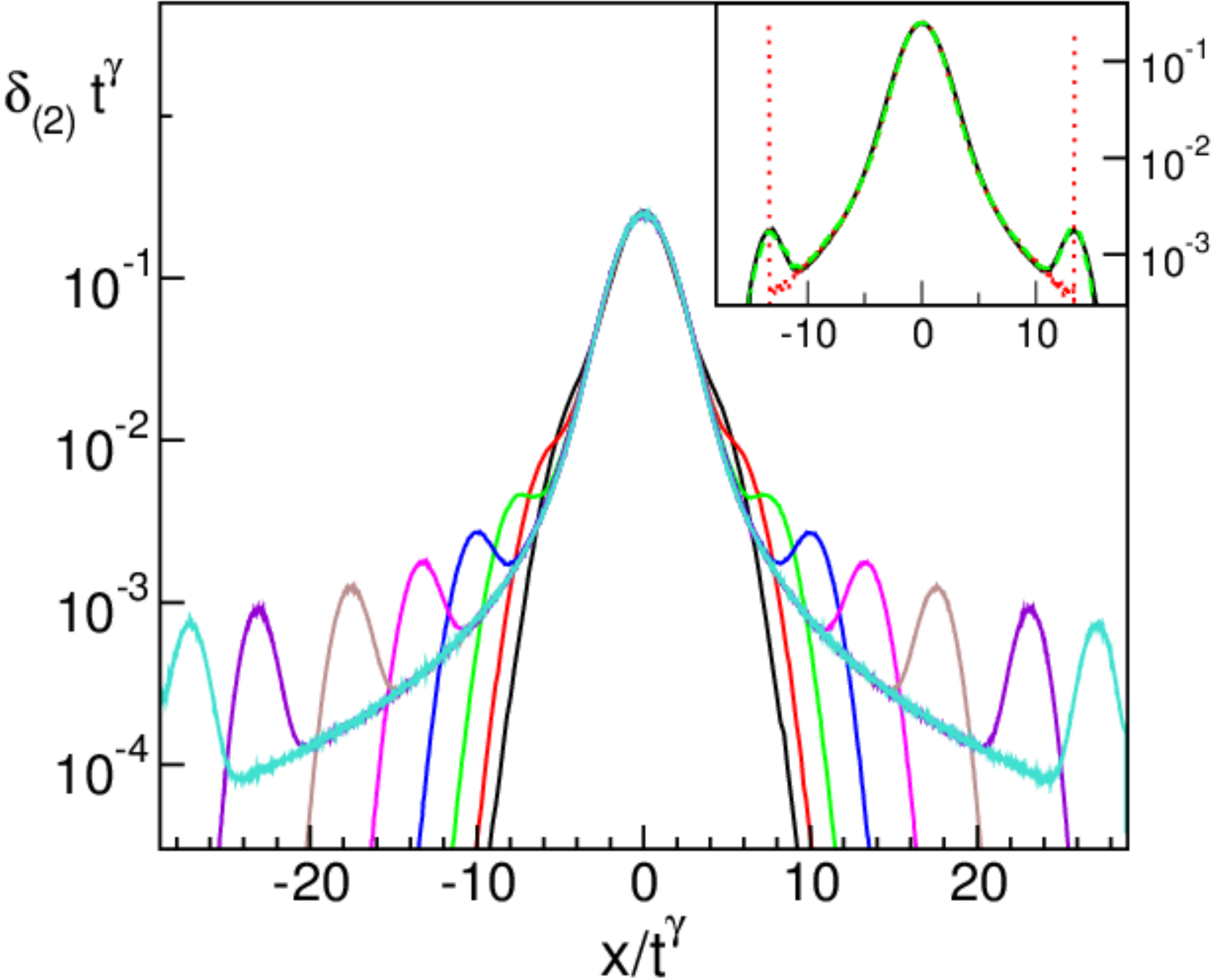}
\caption{Scaled perturbation profiles at times $t=40, 80, 160, 320, 640, 1280, 2560$, and $3840$ , with $\gamma=3/5$. The profiles have been obtained by averaging over large number of  realizations. In the inset, the profile at $t=640$ (solid line) is compared with the propagators of a  L\'{e}vy walk with an exponent $\nu= 5/3$ with a fixed velocity $v= 1$ (dotted line) or with velocity chosen from a  Gaussian distribution with mean $1$ and variance $0.036$ (dashed line).  (Adapted from Cipriani et.al. with permission from  \cite{Cipriani05}  Copyright (2005) by American Physical Society)}
\label{fig:cipriani}
\end{figure} 
\begin{itemize}
\item {\bf Slow decay of energy current auto-correlations}:  A commonly followed approach for determining the $N$ dependence of $j$  or equivalently the thermal conductivity $\kappa$, is to use the {\it closed system Green-Kubo (GK) formula} \cite{Green54, Kubo57}:
\begin{equation}
\kappa=\frac{1}{k_BT^2}\lim\limits_{\tau \to \infty}\lim_{N\to \infty} \frac{1}{N} \int_0^\tau dt~
\langle {J}(t){J}(0) \rangle_{closed}^{eq}~~, \label{GKclosed}
\end{equation}
where ${J}(t)=\sum_x j(x,t)$, with $j(x,t)$ defined in Eq.~(\ref{eq:jmic}), is the total current in the system. The average $\la...\ra_{closed}^{eq}$ is evaluated with initial conditions chosen from a thermal distribution and time-evolution given by the closed system dynamics.    This formula relates the thermal conductivity $\kappa$  to the integral of the equilibrium heat current auto-correlation function $C_J(t)=N^{-1}\langle {J}(t){J}(0) \rangle_{closed}^{eq}$.   Numerical simulations as well as several  theoretical treatments find that  $C_J(t)$ in a closed system generically decays with time  as a power law $C_J(t) \sim t^{\alpha-1}$ with $0 \le \alpha \le 1$ \cite{Lepri03, Dhar08, Lepri98, Delfini07, Lepri98a, Lepri03a, Delfini06, Wang04, Wang04a, Pereverzev03, Lukkarinen08, Basile06, Basile09,Beijeren12, Grassberger02, Casati03, Prosen05,Li04, Wang11, Chen14}.
As an example we show in Fig.~\eqref{fig:grass} data from simulations \cite{Grassberger02} of the alternate mass hard particle gas, where we see a decay with $\alpha \approx 0.33$. Such a power-law time dependence implies, from  Eq.~ \ref{GKclosed}, a divergent thermal conductivity. To see the dependence on  system size one heuristically puts a cutoff  $t_N \sim N$ in the upper limit of the time integral, the argument being that this is the time  taken by sound modes to explore the full system of size $N$.  
Performing the time integral in Eq.~\eqref{GKclosed} with this cut-off, one finally gets $\kappa \sim N^\alpha$. An interesting example where this procedure fails has been pointed out in a recent work \cite{Saito18,Saito18a}.

\item {\bf Super-diffusive spreading of initially localized energy pulse}: Here one  looks at the spreading of a localized energy pulse in a thermally equilibrated system.  One takes an initial configuration chosen from a thermal distribution with  average local energy $e_0=\la e(x) \ra_{\rm closed}^{eq}$, uniform across the system. Imagine putting an extra amount of energy $\epsilon_0$  to a few particles in a region inside the bulk to create a pulse of excess energy locally. As the system evolves according to the closed system dynamics, this localized energy perturbation starts spreading across the system. Let $\epsilon(x,t)$ represent the excess energy density (above $e_0$)  at the point $x$ and at time $t$ (averaged over the initial distribution). This quantity starts as a $\delta$-function at $t=0$ and then starts spreading with time. Note that $\int dx~\epsilon(x,t)=\epsilon_0$, the total injected energy is conserved under the closed system dynamics. For a diffusive system, 
the perturbation would evolve according to the diffusion equation ${\partial  \epsilon (x,t)}/{\partial t} = D {\partial^2 \epsilon(x,t)}/{\partial x^2}$ and in  macroscopic length-time scales, the perturbation profile at time $t$ would be given by a Gaussian
\begin{align}
\epsilon(x,t) = \epsilon_0~\frac{e^{-x^2/4Dt}}{\sqrt{4 \pi D t}}. \label{spreading-diff}
\end{align}
For a system with AHT, one instead finds the following scaling form \cite{Lepri16, Cipriani05}
\begin{align}
 \epsilon(x,t) = \frac{1}{t^{\gamma}}~\mathcal{G}\left(\frac{x}{t^\gamma} \right),~~{\rm for}~~x \lesssim t,   \label{spreading}
\end{align}
with a scaling exponent $1/2<\gamma < 1$. The two limits $\gamma=1/2$ and $1$ correspond respectively to diffusive and ballistic transport. In Fig.~\ref{fig:cipriani} we show results for energy pulse spreading obtained in \cite{Cipriani05} for the alternate mass hard particle gas model.
The main plot shows the scaling $x \sim t^{\gamma}$, with $\gamma=3/5$ of the central part of the distribution. The central part of the distribution was found to fit to the  L\'{e}vy function which is the propagator of Eq.~\eqref{fraceq} with $\mu=1/\gamma$. The mean square deviation (MSD) defined as 
\eqa{
\sigma_e^2(t) = \int dx  ~x^2~ \epsilon(x,t),
}
with mean taken as zero, was seen to scale as $\sigma_e^2(t) \sim t^{\beta}$, with $\beta=4/3$, as opposed to a diffusive system with $\beta = 1$.  It was also noted that the MSD width exponent, $\beta$, is related to the thermal conductivity exponent $\alpha$ as $\beta=1+\alpha$ [see Sec.~(\ref{sec:Levy-open}) for details]. To compute the MSD and relate the exponents $\beta$ and $\gamma$ is a somewhat subtle issue and requires one to note that the scaling function is valid in the bulk region $|x| \lesssim t$, beyond which  $\epsilon(x,t)$  decays rapidly [see discussion in Sec.~\eqref{sec:Levy-closed} in the context of  L\'{e}vy-walk model]. From properties of the   L\'{e}vy distribution  one gets, in the regime $t^\gamma << x \lesssim t$, the scaling form $\mathcal{G}(u) \sim  1/u^{1+1/\gamma}$. Using these asymptotics and  computing $\sigma^2_e(t)=\int_0^t dx ~x^2 ~t^{-\gamma} \mc{G}(x/t^\gamma)$ gives us the leading behavior $\sigma^2(t) \sim t^{3-1/\gamma}$ which then leads to the relation $\beta=3-1/\gamma$.
Observations from  several other numerical simulations have confirmed the super-diffusive behavior \cite{Cipriani05, Li03,Chen13, Wang15, Zaburdaev11, Li15, Liu14}.

\item {\bf Super-diffusive evolution of density correlations}: 
The anomalous signature discussed in the previous point can also be observed alternatively by  looking at the spreading of the equilibrium spatio-temporal correlation function of the energy density $e(x,t)$ defined as 
\begin{align}
C_e(x,t)=\langle e(x,t)e(0,0) \rangle - \langle e(x,t)\rangle \langle e(0,0) \rangle, \label{e-corr-closed}
\end{align}
where the average is taken over the equilibrium initial conditions. For diffusive systems this correlation has the Gaussian form in Eq.~\eqref{spreading-diff}, while for systems with AHT this has the scaling form in Eq.~\eqref{spreading} and one again has super-diffusive growth of the MSD \cite{Zhao06}, now defined as \eqa{
\sigma_c^2(t)=\frac{1}{k_B T} \int dx ~x^2~ C_e(x,t).
}
This MSD can be related to $\sigma_e^2(t)$ defined above, using linear response theory and both have $\sim t^\beta$ scaling.  
In the  case of AHT, observing the scaling form in Eq.~\eqref{spreading} usually requires one to subtract contributions of sound modes which travel ballistically. The theory of nonlinear fluctuating hydrodynamics (NFH) provides a framework in which one can systematically describe the super-diffusive scaling of the correlation \cite{Narayan02,Beijeren12,Spohn14, Spohn15, Mendl13, Das14}. This theory is based on writing hydrodynamic equations for the conserved quantities in the system which for the Hamiltonian in Eq.~\eqref{hamiltonian} are the total energy, total momentum and the total number of particles (or volume). This framework of NFH is discussed in detail in Sec.~\eqref{sec:NFH}.
A connection can be made between the super-diffusive scaling ($\sigma_c^2(t) \sim t^\beta$)  of the energy correlations and  the power-law decay, $\sim t^{\alpha-1}$, of the current-current correlations \cite{Liu14,Li15,Lepri16}, which can be seen as follows.  Starting from  the continuity equation for energy, one can obtain the relation \cite{Spohn14,Spohn15} on the infinite line
\eqa{
\frac{\p^2 C_e(x,t)}{\p t^2}=\frac{\p^2 \la j(x,t) j(0,0) \ra}{\p x^2}~.
}
Multiplying by $x^2$ on both sides and integrating over all the range of $x$ one gets
\eqa{
\frac{d^2 \sigma_c^2(t)}{dt^2}= \frac{1}{k_B T} \la J(t) j(0,0)\ra = \frac{C_J(t)}{k_B T}~. 
}
Assuming the expected forms $\sigma^2(t)\sim t^\beta$ and $C_J(t) \sim t^{\alpha-1}$ we get the relation
$\alpha=\beta-1$.

\end{itemize}

\begin{figure}
\includegraphics[width=0.75\linewidth]{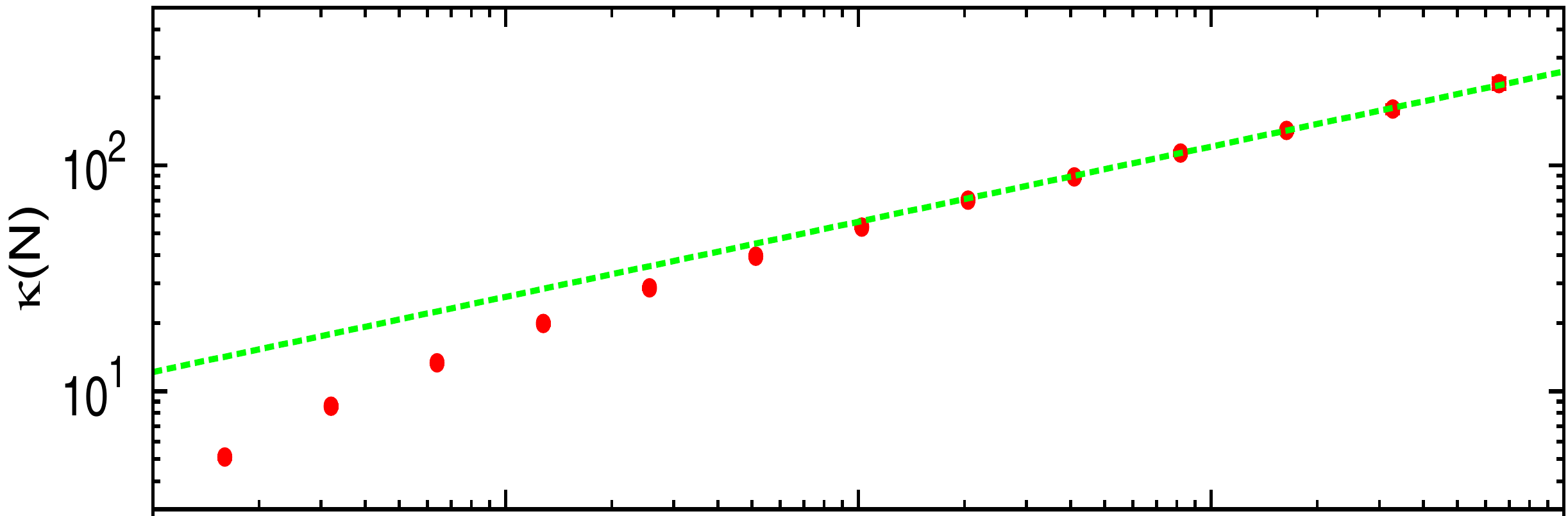}
\caption{FPUT-$\beta$ model: Results for conductivity $\kappa$ versus $N$ for $T_\ell=2.0$ and $T_r=0.5$. The last five points fit to a slope of $0.333\pm 0.004$
	(Adapted from Mai et.al. with permission from \cite{Mai07} Copyright (2007) by American Physical Society) }
\label{fig:fpukappa}
\end{figure} 

\begin{figure}
\includegraphics[width=0.75\linewidth,angle=0]{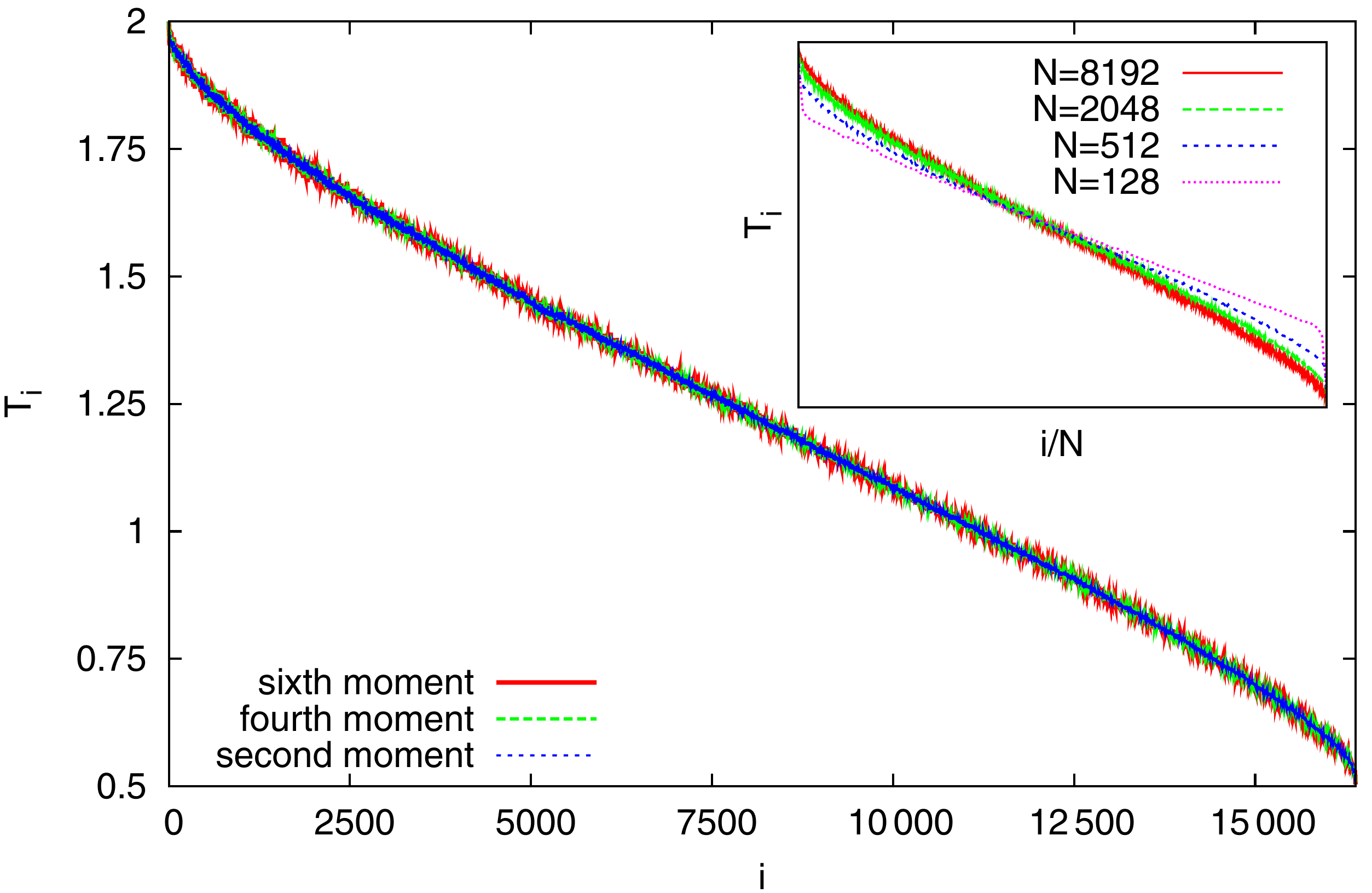}
\caption{FPUT-$\beta$ model: Kinetic temperature profile for a system  with $N=16384$, $T_\ell=2.0, T_r=0.5$. Assuming a Gaussian local velocity distribution, the temperatures as defined from the first three even moments  are  shown; their agreement vindicates the assumed Gaussian velocity distribution. (Inset) Normalized temperature profiles for different $N$. 	(Adapted from Mai et.al. with permission from \cite{Mai07} Copyright (2007) by American Physical Society)}
\label{fig:fputemp}
\end{figure}

\subsection{Signatures in the open system set-up}
\label{sec:sign-open}
\begin{itemize}
\item {\bf Diverging thermal conductivity}: As discussed above, in the open system set-up, one connects the system at the two boundaries to  heat reservoirs at  
unequal temperatures $T_\ell \neq T_r$.  A common  model for baths is to write  Langevin dynamics for the boundary particles involving  dissipation and noise term satisfying the fluctuation-dissipation relation. For a chain of interacting particles  described by the Hamiltonian in Eq.~\eqref{hamiltonian} the  equations of motion for the boundary particles would read 
\begin{align}
\dot{p}_1&=f_1- {\lambda} p_1 +  \xi_\ell(t), \\
\dot{p}_N&=f_N- {\lambda}p_N +  \xi_r(t),
\label{Langevin-eqs}
\end{align}
where $f_i=-{\partial H}/{\partial q_i}$. The noise terms $\xi_{\ell,r}$ are Gaussian white noise terms, with zero mean and correlations $\la \xi_\ell(t) \xi_\ell(t')=2 \lambda T_\ell \delta(t-t')$ and $\la \xi_r(t) \xi_r(t')=2 \lambda T_r \delta(t-t')$. The remaining particles evolve according to Eq.~(\ref{eqmot}).
 After a long time the system reaches a non-equilibrium steady state (NESS) 
and we can measure the steady state current $j$ as average of the local current $j(x,t)$ defined through Eq.~\eqref{eq:jmic}.  In the steady state this will be independent of time as well as the bond where we measure the current. 
One can then check if the system size $N$ scaling of this steady state current $j$ has the expected form  $j \sim {N^{\alpha-1}}$, where $\alpha <1$ for anomalous systems. Alternatively one can define the 
$\kappa = {j N}/{(T_\ell-T_r)}$ and see how this scales with $N$. For a large class of nonlinear interaction potentials, it has been observed that the thermal conductivity $\kappa \sim N^\alpha$ with $0<\alpha<1$ for large $N$ \cite{Dhar01, Grassberger02, Mai07, Dhar08a, Das14, Deutsch03}. As an example, we show in Fig.~\eqref{fig:fpukappa} data from \cite{Mai07} for the FPUT-$\beta$ chain, where one finds $\alpha \approx {0.33}$. 

\item {\bf Non-linear temperature profile}: The local temperature at a site on the lattice can be defined through the relation $T_i=\la p_i^2/m\ra$, where the average is taken in the NESS. For diffusive systems, the temperature profile obtained would be linear for small $\Delta T=T_\ell-T_r$, as expected from solving Fourier's law with a constant $\kappa$.  It is important to note that non-linear temperature profiles can also be obtained in case of diffusive transport if the thermal conductivity $\kappa$ is temperature-dependent and $\Delta T$ is large.
On the other hand, for many systems with AHT,  one finds a strongly non-linear temperature profile even when $\Delta T$ is made arbitrary small \cite{Lepri97,Mai07,Roy12,Dhar08a,Lepri09,Cividini17,Kundu18}.   Quite often the profiles are characterized by  divergent slopes at the boundaries.
In Fig.~(\ref{fig:fputemp}) we show  the temperature profile in the FPUT-$\beta$ model and one can see the characteristic non-linear nature.  
Note that the definition of local temperature makes sense (and is useful) only if this temperature predicts correctly other local observables, for example higher moments of the velocity. This was also verified in \cite{Mai07} and also shown in Fig.~\eqref{fig:fputemp}.  
Typically one finds that the temperature difference $\delta T(x)=|T_\ell-T(x)|$ scales as $(\delta x)^\mu$, with distance $\delta x$ from the boundary, where $0< \mu \leq 1$. The exponent $\mu$ has been referred to as the meniscus exponent \cite{Lepri11}. This exponent is non-universal in the sense that it depends on details of boundary conditions, unlike the conductivity exponent $\alpha$.

\item {\bf Green-Kubo-type relation for open systems}:  Analogous to the Green-Kubo formula in the closed system set-up given by Eq.~\eqref{GKclosed}, an exact   formula exists in the open system set-up that relates the current response to a small temperature  difference $\Delta T=T_\ell-T_r$. This is given by  \cite{Kundu09} 
\begin{equation}
\lim_{\Delta T \to 0} \frac{j}{\Delta T} = \frac{1}{K_BT^2 N^2} \int_0^\infty dt~
\langle {J}(t){J}(0) \rangle_{open}^{eq}~. \label{GK-open}
\end{equation}
 The  time auto-correlation $\langle {J}(t){J}(0) \rangle_{open}^{eq}$ is computed by averaging over equilibrium initial conditions as well as the open system dynamics which includes  the stochastic baths (at equal temperatures). This formula is valid  for a finite size system.  We note that for systems with AHT, unlike with Eq.~\eqref{GKclosed}, in the open set-up we do not require the use of an
 upper cut-off $t_N \sim N$ for estimating the size dependence of conductivity.  In this case the linear response current can be evaluated directly from Eq.~\eqref{GK-open} for any finite system of size $N$ and thereby one can verify the form $j/\Delta T \sim N^{\alpha -1}$. This approach has been discussed for example in \cite{Deutsch03,Das14}. It was observed in \cite{Deutsch03} that, for the so-called random collision model studied by them,   both  $\langle {J}(t){J}(0) \rangle_{closed}^{eq}$ and $\langle {J}(t){J}(0) \rangle_{open}^{eq}$ showed a $t^{-2/3}$ decay at times $t \lesssim N$. However  the exponential decay for the open case begins at $t_N \sim N$ while for the closed system (with periodic boundary conditions) this begins at $t_N \sim N^{3/2}$. This was understood as arising from the time scale associated to the spreading of sound modes. Note that if we put the cut-off $t_N \sim N^{3/2}$ as the upper limit in the time-integral of Eq.~\eqref{GKclosed} then we would get the wrong conductivity exponent. In order to get the correct exponent in the closed system set-up, one has to by hand set the cut-off at $t_N\sim N$ based on consideration of the practical transport set-up  which has baths at the boundaries.

Recently, in a model system of AHT the  relation in Eq.~\ref{NLFL} has been established using the above formula and a heuristic approach based on fluctuating hydrodynamics \cite{Cividini17}.  An explicit expression of the  kernel was obtained for a specific model, using which one can understand the divergence of $\kappa$ as well as the  singular features in the temperature profile . A detailed discussion of this method is given later in Sec.~(\ref{sec:BC}).

\end{itemize}


\section{Phenomenological approaches for anomalous heat transport}
\label{sec:theory}
In this section we will discuss two different approaches that have tried to understand the various aspects of AHT mentioned above. The first is a completely heuristic approach where one assumes that the heat carriers perform  L\'{e}vy walks instead of random walk which is expected for diffusive heat transfer. This method has been used to explain spreading of localized energy pulses, steady state properties and current fluctuations \cite{Denisov03,Cipriani05,Delfini07,Lepri11,Zaburdaev11,Dhar13,Dhar13b,Dhar18}. 
The second  approach is a microscopic one where one starts by writing hydrodynamic equations for the conserved quantities of the Hamiltonian dynamics. One then phenomenologically adds noise and dissipation terms satisfying fluctuation dissipation relations and this allows one to study equilibrium fluctuations in the system. In particular, using the formalism of fluctuating hydrodynamics, one can compute dynamical correlation functions which contain information on AHT. 

\subsection{ L\'{e}vy walk description of anomalous heat transport}
\label{sec:Levy}

\subsubsection{ L\'{e}vy walk description in the closed set-up}
\label{sec:Levy-closed}
In this description one thinks of  energy   as being  carried by  L\'{e}vy walkers, 
each of which carry  a fixed amount of energy. It follows that the local energy density and energy current at  any point  can be taken to be directly  proportional to, respectively,  the particle density and current. Let us also assume that the local temperature is proportional to the local energy density and hence to the  density of particles.

{\bf Definition of the  L\'{e}vy walk} \cite{klafter93,zumofen93,metzler99}:  At each step of the walk, a particle chooses a time of 
flight  $\tau$ from  a specified  distribution,  $\phi(\tau)$, and then moves    a distance  $x=c \tau$ at  a fixed speed $c$, with equal probability in either direction. More generally one can consider the velocity $c$ to be drawn from a distribution. Let $P(x,t) dx$ denote  the probability 
that the particle is in the interval  $(x,x+dx)$ at time $t$. Note that $P(x,t)$  also includes events where the particle is crossing the interval $(x,x+dx)$, in addition to the events in which the particle lands in the interval at time $t$. 
If a particle starts at the  origin at time $t=0$, the probability  $P(x,t)$ satisfies
\begin{align}\label{P(x,t)}
 P(x,t)&= {1 \over 2}\psi(t) \delta(|x|- c t )  
+ {1 \over 2} \int_0^t   d \tau \phi(\tau)  [P(x- c \tau,t-\tau) + P(x+ c \tau,t-\tau) ]~,
\end{align}
where  $\psi(\tau)=\int_\tau^\infty d\tau'~\phi(\tau')$ is the probability of choosing a time of flight $\geq \tau$. Here we   consider  L\'{e}vy walkers with a time-of-flight distribution 
\bea
~\phi(t) &=& \f{\nu}{t_o}\f{1}{(1+t/t_o)^{\nu+1}}~,~~1<\nu<2~. \label{phiform}
\eea
which decays, at large times,  like a power law  $ \phi(t) \simeq A ~t^{-\nu-1} $ with $A=\nu t_0^\nu$.
 For  this  range of $\nu$  the mean flight time 
 $\la t \ra =\int_0^\infty dt ~t ~ \phi(t)= t_0/(\nu-1) $ is finite but   $\langle t^2 \rangle = \infty$.

{\bf Some properties of the  L\'{e}vy walk}: Taking the Fourier Laplace transform $\widetilde{P}(k,s)=\int_{-\infty}^\infty dx \int_0^\infty ~dt ~P(x,t) ~e^{i k x-st}$ we get
\bea
\widetilde{P}(k,s)=\f{\widetilde{\psi}(s-ick)+\widetilde{\psi}(s+i ck)}{2-\widetilde{\phi}(s-i ck)-\widetilde{\phi}(s+i ck)}~,
\eea 
where $\widetilde{\phi}(s)=\int_0^\infty dt e^{-st} \phi(t)$ and 
$\widetilde{\psi}(s) =\int_0^\infty dt e^{-st} \psi(t) =[1-\widetilde{\phi}(s)]/s$.

\begin{figure}
\includegraphics[width=0.75\linewidth]{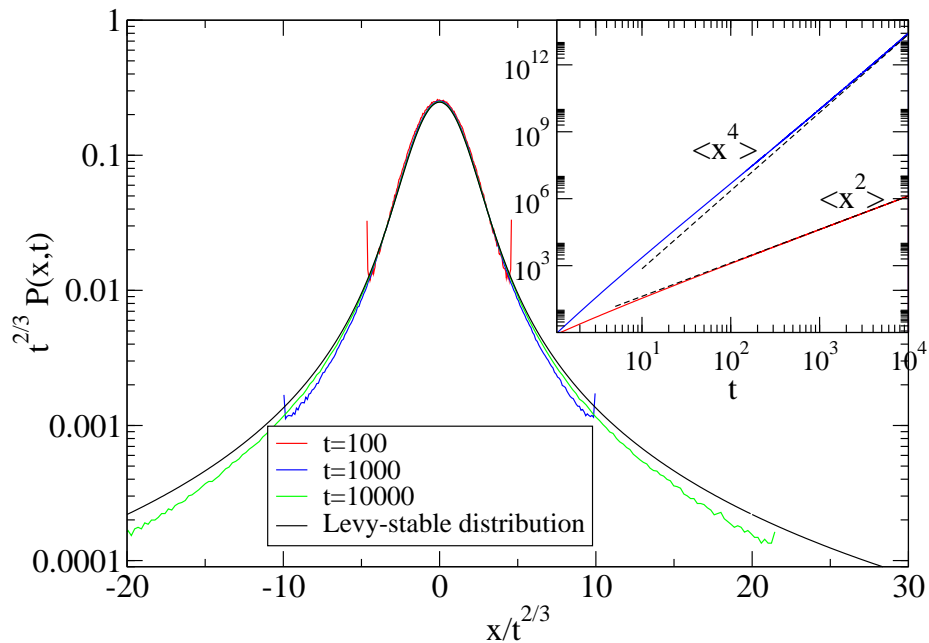}
\caption{Plot of the scaled distribution $t^{2/3}P(x,t)$ versus $x/t^{2/3}$ of the  L\'{e}vy walk on the open line for $\nu = 3/2$ at three different times. Also shown is a plot of the  L\'{e}vy-stable distribution. The inset shows a plot of the mean square displacement and the fourth moment and a comparison with the exact asymptotic forms (dashed lines) given by Eqs.~(\ref{xsq},\ref{xqd}).  In all plots the time $t_o$ and $c$ are set to one.
 }
\label{fig2}
\end{figure}

For asymptotic properties it is useful to find the form of $\widetilde{P}(k,s)$ for small $k,s$. The Laplace transform $\widetilde{\phi}$ is given by:
\bea
\widetilde{\phi}(s) &=&\int_0^\infty dt~e^{-st}~\phi(t)=1-\la t \ra ~s+
b~\nu (s t_o)^\nu+\cdots ~,  \\
{\rm where}~~b &=& \f{1}{\nu (\nu-1)}~\int_0^\infty~ dz~e^{-z} z^{1-\nu}= \f{1}{\nu(\nu-1)} \Gamma(2-\nu)~, \nn 
\eea
and $\Gamma(u)$ is the Gamma-function.  Hence we get:
\bea
\widetilde{P}(k,s)= \f{1  - d [(s-ick)^{\nu-1} +(s+ick)^{\nu-1}]}
{s  - d [(s-ick)^{\nu} +(s+ick)^{\nu}]}~, \label{propks}
\eea
where $d=b A/(2 \la t\ra)$.
Taking the inverse Fourier-Laplace transform of this gives us the propagator 
of the  L\'{e}vy walk on the infinite line.  This   
 corresponds to a pulse whose central region is a  L\'{e}vy-stable distribution with a scaling $ x \sim t^{1/\nu}$. This can be seen by expanding Eq.~(\ref{propks}) for $ck/s <<1$ to get $\widetilde{P}(k,s)=[s-c \cos (\nu \pi/2) (ck)^\nu]^{-1}$.  The difference with the  L\'{e}vy-stable distribution is that the  L\'{e}vy-walk propagator has ballistic peaks  of  magnitude $t^{1-\nu}$ at $x=\pm ct$  and vanishes outside this.  The overall behavior of the propagator is as follows \cite{klafter93}: 
\bea \label{Levyprop}
P(x,t) &\sim t^{-1/\nu} ~\exp \left(\frac{-a x^2}{t^{2/\nu}}\right) ~~~~|x| \lesssim t^{1/\nu} \nn \\
&\sim t~x^{-\nu-1}~~~~~~~~~~ t^{1/\nu} \lesssim |x| < ct \nn \\
&\sim t^{1-\nu} ~~~~~~~~~~~~~~~~|x|= c t \nn \\
&= 0~~~~~~~~~~~~~~~~~~~~ |x| > ct ~.
\eea
The time evolution of the  L\'{e}vy-walk propagator, obtained from direct simulations of the  L\'{e}vy walk, is shown in 
Fig.~(\ref{fig2}). We also plot the  L\'{e}vy-stable distribution obtained by taking the Fourier transform of $P(k,t)=e^{-c \cos (\nu \pi/2) |k|^{\nu}t}$.

Various  moments of the distribution can be found using the relation 
$\la x^n \ra(t)=  ({d^n}/{d (ik)^n}) \int_{-\infty}^\infty dk e^{ikx} P(x,t) \Big|_{k=0}~$, or its Laplace transform  given by $ \la x^n \ra(s) = ({d^n}/{d (ik)^n}) ~ \widetilde{P}(k,s)~\Big|_{k=0}$.
Using Eq.~(\ref{propks}) we get in particular the following leading behavior
\begin{align}
\langle x^2 \rangle_{{c}}  &\simeq    {2~ A ~ c^2\over (3-\nu)(2-\nu) \nu   ~ \langle \tau \rangle} \,  t^\beta,~~~~\beta={3-\nu}~, \label{xsq}   \\ 
\langle x^4 \rangle_c  
&  \simeq    
{  
{4~ A ~ c^4 \over (5-\nu)(4-\nu) \nu   ~ \langle \tau \rangle} \,  t^{\beta+2} ~.
} 
 \label{xqd}
\end{align} 
 We see that for $1<\nu<2$  the motion is  super-diffusive  \cite{zumofen93,metzler99}. The most interesting characteristics to note about the  L\'{e}vy walk are 
the fact that the probability distribution has finite support  ($|x| \leq ct$), in the bulk it coincides with the  L\'{e}vy distribution with scaling $x \sim t^{1/\nu}$ and  finally the mean square displacement (MSD) $ \la x^2 \ra \sim t^{\beta}$ with $\beta=3-\nu$. Note that the usual  L\'{e}vy stable distribution has a diverging second moment, however the  L\'{e}vy walk has a finite MSD and this follows from the finite support $|x| \leq ct$ of the corresponding distribution. Indeed, on using this cutoff and the power-law form of the  L\'{e}vy near the cut-off (see Eq.~\eqref{Levyprop}) gives us the expected scaling exponent $\beta=3-\nu$.

{\bf  L\'{e}vy walks and AHT}: The first proposal suggesting the  L\'{e}vy walk model to describe anomalous heat transport was made in \cite{Denisov03}. This idea was tested for a microscopic model in \cite{Cipriani05} where it was shown that the spreading of a heat pulse in a thermally prepared  alternate mass hard particle gas  was super-diffusive and is well-described by the  L\'{e}vy walk model. In Fig.~\eqref{fig:cipriani} we show the evolution of a localized perturbation. The main plot shows the $x \sim t^{3/5}$ scaling of the central part of the distribution while the inset shows a fit to the expected  L\'{e}vy distribution (for a LW with $\nu=5/3$) with a single fitting parameter. It was also shown that the MSD of the energy has the scaling $\sim t^{4/3}$ as expected from the relation $\beta= 3-\nu$ for LW. Finally it was proposed using linear response ideas that the exponent $\beta$ and the conductivity exponent $\alpha$ should be related as $\alpha=\beta-1$ which gives $\alpha=1/3$ for the present system. This agrees with known results for the alternate mass hard particle gas. The validity of the  L\'{e}vy walk description of pulse propagation was further verified in \cite{Delfini07} for a hard particle gas interacting via a square well potential and in \cite{Zaburdaev11} for the FPUT chain. All these cases were described by the same  L\'{e}vy-walk exponent $\nu=5/3$.

\subsubsection{ L\'{e}vy walk description of the open set-up} 
\label{sec:Levy-open}
\begin{figure}
 \includegraphics[width=0.85\linewidth]{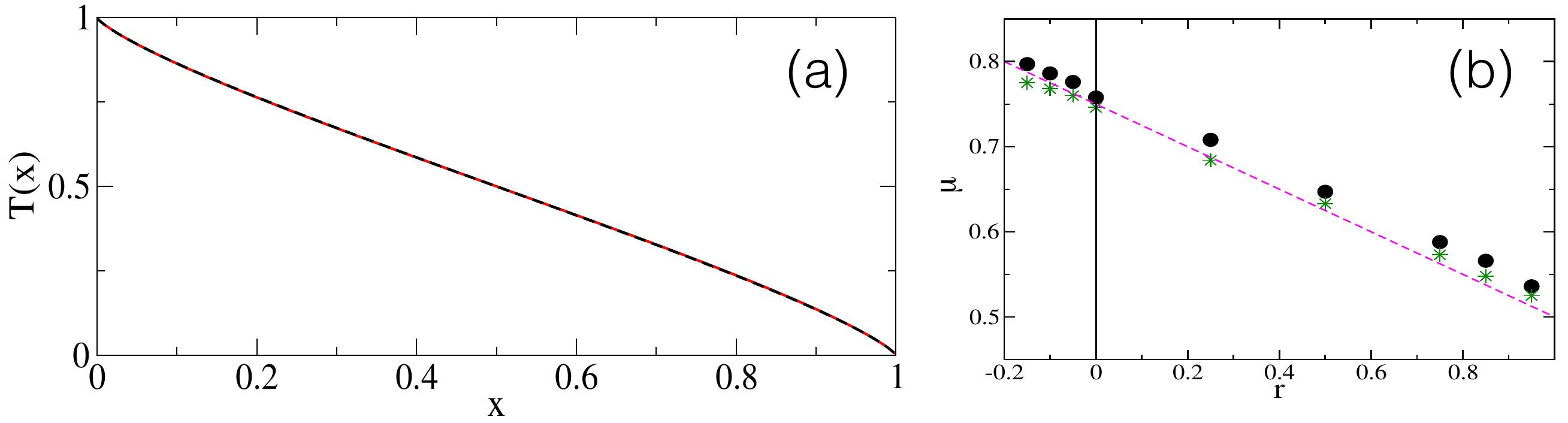}
\caption{{\bf (a)} Temperature profile of the oscillator chain with conservative noise with free boundary condition and $\lambda=\gamma=1$ (solid line) and solution of the master equation with reflection coefficient $R=-0.1$ (dashed line). (Reprinted  from \cite{Lepri11}). {\bf (b)} Dependence of the meniscus exponent $\mu$ on the reflection probability $r$ for $\nu=3/2$. Note that $r$  in this figure is denoted by $R$ in this review. Full circles and stars are measures from fitting of $P(x)$ and $Q(x)$, respectively (see text). The dashed line is given by formula \eqref{Levymueq}. 	(Adapted from Lepri  et.al. with permission from \cite{Lepri11} Copyright (2011) by American Physical Society).}
 \label{levyR0prof}
 \end{figure}

We now discuss the case of the open system consisting of a finite segment $(0,L)$  that is connected to two reservoirs at the ends. 
The use of the  L\'{e}vy walk model to study NESS properties in AHT was first proposed in \cite{Lepri11} where the authors considered a finite lattice of $N$ sites containing a collection of  L\'{e}vy walkers. The system was connected at it's two ends to infinite reservoirs that contained sources  emitting  L\'{e}vy walkers at fixed constant rates. A  L\'{e}vy walker crosses from the reservoir into  system with probability one, but while exiting from system into reservoir, it can get reflected with probability $R$. A particle exiting the reservoir is eliminated.
The authors in \cite{Lepri11} considered a discrete version and studied this problem numerically. The strategy was to write appropriate master equations for the probability evolution and obtain the steady state solution numerically. One of the main observations in the paper was that the NESS profile for $P(x)$ was non-linear and was singular at the boundaries. In Fig.~\eqref{levyR0prof}(a) we show a plot of the profile for the case $R=-0.1$, compared with simulation results for the temperature profile in the momentum exchange model (HCME), with free BC and a specific choice of exchange rate. One sees very good agreement. As noted in Sec.~(\ref{sec:sign-open}) in the context of temperature profiles in systems with AHT, one can define a ``meniscus'' exponent, $\mu$,  through the observed scaling form $P(x)\sim (\delta x)^\mu$ for small distances $\delta x$ from any boundary.  Based on their numerical observations [see Fig.~\eqref{levyR0prof}(b)] the authors in \cite{Lepri11} conjecture the relation 
\eqa{\label{Levymueq}
\mu=\frac{\nu}{2}+R\left(\frac{\nu}{2}-1\right)~.
}
It was noted in \cite{Lepri11} that the value $R=-0.1$ was unphysical  but made 
mathematical sense in the master equation (see \cite{Lepri11}  for further discussions on this point) and gave the best agreement with the momentum exchange simulation profile.

%

Some exact results were obtained for the  L\'{e}vy walk model with particle reservoirs, for the special case of perfectly transmitting boundary walls (i.e $R=0$) \cite{Dhar13} which  we now describe. We note that for the  L\'{e}vy walker, at any given time, a particle could either be   passing over a  point $x$ or could have landed precisely at the point. Hence, in addition to the probability density $P(x,t)$, it is convenient to define the probability $Q(x,t) dx dt$ that a particle lands precisely between  $x$ to $x+dx$ in the time interval $(t,t+dt)$. 
We now specify the   boundary conditions required to set up 
a non-equilibrium current carrying  steady state. 
For this, we consider the region $  x \le 0$  as the left reservoir with $Q(x,t)=\ql$ for all points in this region. Similarly,  we set $Q(x,t)=\qr$  in the region $x \geq L $ corresponding to the right reservoir. In the steady state, the distributions become time-independent and $Q(x,t)=Q(x)$, $P(x,t)=P(x)$  satisfy \cite{Dhar13}
 \begin{align}
& Q(x)=\int_0^L  dy~\f{1}{2c} \phi(|x-y|/c) ~Q(y) +  \f{\ql}{2} \psi(x/c)
+ \f{\qr}{2} \psi[(L-x)/c]~,  \label{QeqSS} \\ 
& P(x)= \int_0^L  dy \f{1}{2c} \psi(|x-y|/c)~ Q(y) +  \f{\ql}{2} \chi(x/c) + \f{\qr}{2} \chi[(L-x)/c] ~. \label{PeqSS}
\end{align}
 where $\psi(t)=\int_t^\infty d\tau ~\phi(\tau )$ and $\chi(t ) =  \int_{t}^\infty d \tau \psi(\tau)$ as mentioned earlier after Eq.~(\ref{P(x,t)}).  The terms on the right hand side  of the above equation for $Q(x)$  represent different contributions. The first term represents the contributions from walkers that start from 
various points $y$ and land  at $x$. The second and the third term represent contributions from walkers starting, respectively, from left and right reservoirs and landing at $x$.  Similarly, the terms on the right hand side of Eq.~\eqref{PeqSS} for $P(x)$ can also be interpreted in the same way except now,  events in which the walkers are passing over $x$, in addition to the events in which they land at $x$ at a given time, also contribute.

Interestingly, it turns out that the problem of finding $Q(x)$ can be related to the problem of the escape probability \cite{Buldyrev01} of a  L\'{e}vy walker on the interval $(0,L)$. Let  
 $H(x)$ denote the probability with which a  L\'{e}vy walker, starting at position $x$, arrives at the left reservoir  (region $x < 0$) before arriving at the right reservoir (region $x >L$).  It can be shown that $ H(x)$ satisfies \cite{Dhar13}
\begin{align}
 H(x)=\int_0^L  dy~\f{1}{2c} \phi(|x-y|/c) ~H(y) 
+ \f{1}{2} \psi(x/c)
~.  \label{H(x)} 
\end{align}
The probability $Q(x)$ can now be expressed in terms of  $H(x)$ as $Q(x) = (\ql-\qr)  H(x) + \qr$, which can be checked easily to satisfy Eq.~(\ref{QeqSS}).
 
If one considers a  L\'{e}vy flight with distribution   $\rho(z)=[\phi(z/c)+\phi(-z/c]/(2 c)$  of steps $z$, the probability $H(x)$   that  starting at $x$, the  flight hits first the left 
bath satisfies exactly Eq.~(\ref{H(x)}).  Hence
 by following the same mathematical steps as in     \cite{Buldyrev01} to study equations such as (\ref{QeqSS}) or (\ref{H(x)}), one  can show  that, in the large $L$ limit,  the solution  $Q(x)$ of (\ref{QeqSS})   (and $H(x)$ of (\ref{H(x)}))  satisfies
\begin{align}
\int_0^L dy~\psi (|x-y|/c)~ {\rm Sgn}(x-y) Q'(y)= 0~.  \label{Qpeq}
\end{align}
 with $Q(0)=\ql$ and $Q(L)=\qr$ [and $H(0)=1$ and $H(1)=0$ for  (\ref{H(x)})] with a  solution of (\ref{Qpeq}), for  a $\phi(\tau)$    decaying as in (\ref{phiform}), which satisfies 
\begin{align} 
Q'(x)=-B [x (L-x)]^{\nu/2-1}~. \label{Qpsol}
\end{align}
We can integrate this equation to get $Q(x)$, with the integration constant  and $B$ being then determined from  the boundary conditions $Q(0)=\ql$ and 
$Q(L)=\qr$.  One finally obtains 
\begin{align}
Q(x)&=Q_l+(Q_r-Q_l)~\mathbb{G}\left ( \frac{x}{L}\right),~~\text{where,}~~
\mathbb{G}(z)=\frac{2 \Gamma(\nu)}{\nu \Gamma(\nu/2)^2}~z^{\nu/2}~_2F_1\left(1-\frac{\nu}{2},\frac{\nu}{2},1+\frac{\nu}{2},z \right),
\end{align}
where $_2F_1\left(a,b,c,z \right)$ is the hypergeometric function. 
For large $L$,  the right hand side of Eq.~(\ref{PeqSS}) is dominated by the range $| y-x|\ll L$ and therefore
\begin{align}
P(x)=\chi(0)Q(x) =  \langle \tau \rangle Q(x) ~. \label{pq}
\end{align} 
The exact results of  Eqs.~(\ref{Qpsol}) have been verified in \cite{Dhar13} from direct  numerical solution of Eqs.~(\ref{QeqSS},\ref{PeqSS}) and it was noted that density profiles were similar to the temperature profiles seen in AHT.

Next we discuss the  steady state current $j(x)$ which is given by 
\begin{equation}
 j(x) = \frac{1}{2}\int_{-\infty}^{\infty} dy  ~ Q(x-y) 
 ~{\rm Sgn} (y) ~ \psi(|y|/c) ~. \label{currexp}
\end{equation}
This can be seen to be  the difference  between  the flow from left to right and from right to left.
 The contribution  from $0<y<\infty$ to the integral corresponds to 
particles crossing the point $x$ from left to right which is obtained by multiplying the density of particles at $x-y$  with the probability $\psi(y/c)$  that they have a flight time larger than $y/c$. The contribution from 
$-\infty < y <0$ to the integral corresponds to a similar right-to-left current. 
After performing a 
partial integration  and using the boundary conditions 
$Q(0) = \ql$ and $Q(L) = \qr$, one obtains
\begin{align}
j(x)&= -\f{c}{2} \int_{0}^{L} dy~ \chi (|x-y| / c)~ Q'(y)~. \label{jss}
\end{align}
Using Eq.~(\ref{Qpeq}) it is easy to see that $dj/dx=0$ which implies that the current in the steady state is independent of $x$, as  expected.  Hence,  evaluating the current at $x=0$ and  using  
Eq.~(\ref{Qpsol}), we get for large $L$ 
\begin{align}
j \simeq   ( \ql- \qr)  \  
{ A~ c^{\nu} 
~\Gamma(\nu )~ \Gamma(1- {\nu \over 2}) \over 2 ~\nu (\nu-1)~\Gamma({\nu \over 2}) 
}   \  L^{\alpha -1}  ,~~~\alpha= 2-\nu.
 \label{Jsol}
\end{align}
From Eq.~(\ref{xsq}) we then get the relation $\alpha=\beta-1$, 
 between the conductivity exponent 
of AHT and the MSD exponent for 
 L\'{e}vy-walk diffusion. This relation for  L\'{e}vy diffusion was pointed out in  
 \cite{Denisov03} and verified in simulations in $1D$ heat conduction models \cite{Cipriani05,Zhao06}. A derivation based on linear response theory has  been given in  \cite{Liu14}. Finite size corrections to the results in Eqs.~(\ref{Qpsol},\ref{Jsol}) were recently obtained in \cite{Miron19a}.

In the  large $L$ limit by using Eq.~(\ref{pq}) in Eq.~(\ref{jss}) 
we obtain 
\begin{align}
j&=-\f{c}{2 \langle  \tau \rangle }\int_0^L d{ y}~ \chi(|x - y| /c) P'(y )~. 
\end{align}
Above equation is the analogue of Fourier's law Eq.~(\ref{FL}) with the important difference that in the linear response regime the current at a point gets contributions from the temperature gradients at other parts of the 
system as well.

The above treatment can be generalized for arbitrary values of the reflection probability $R$ \cite{Kundu19} and this leads to the following general non-local form of the  current
\eqa{
j&= -\f{c}{2 \langle  \tau \rangle }\int_0^L d{ y}~ \chi_R(x,y) P'(y )~,\\
{\rm where}~~\chi_R(x,y) &=\sum_{n=-\infty}^\infty \left[ R^{|2n|}\chi \left(\frac{|2nL+y-x|}{c}\right)-  R^{|2n+1|}\chi \left(\frac{|2nL+y+x|}{c}\right) \right]~. \label{LWchiR}
}
Remarkably we note that for $\nu=3/2$ ($\alpha=2-\nu=1/2$), the expression above is identical to the expression for $\mc{K}_R(v,v')$  with $v=x/L,v'=y/L$, given later in  Eq.~\eqref{eq:solK}.
\subsection{Nonlinear fluctuating hydrodynamics description of anomalous heat transport}
\label{sec:NFH}
We now discuss a completely different approach for understanding AHT.  In this approach the starting point is the Hamiltonian dynamics of the system. The idea is to  consider the effective dynamics of the slow conserved fields  using some coarse graining. One finds that the evolution of small fluctuations around equilibrium can be described by fluctuating hydrodynamics \cite{Narayan02,Beijeren12,Spohn14, Spohn15, Mendl13}.
Solving these equations using mode coupling theory,  detailed predictions can be made on the form of equilibrium spatio-temporal correlation functions of the conserved fields. In particular, we will see that it predicts the super-diffusive spreading of energy perturbations with  L\'{e}vy-law scaling, and the slow decay of energy current auto-correlation functions. We will here describe the theory for generic anharmonic systems with three conserved quantities, namely volume, momentum, energy \cite{Spohn14} and present some numerical results which verify the predictions of the theory.

 Let us consider $N$ particles of unit masses with positions and momenta denoted by $\{q(\ell),p(\ell)\}$, for $\ell=1,\ldots,N$. The particles move on a ring of size $L$  so that we have the boundary conditions $q({N+1})=q(1)+L$ and $p({N+1})=p(1)$. 
The Hamiltonian   is taken to be 
\bea  \label{ham}
H = \sum_{\ell=1}^N \epsilon(\ell), \quad
\epsilon(\ell) = \f{p^2(\ell)}{2}+ V[r(\ell)]~, 
\eea 
where we have defined the stretch variables $r(\ell)=q(\ell+1)-q(\ell)$. It is easy to see from the Hamiltonian equations of motion that stretch $r(\ell)$, 
momentum $p(\ell)$, and energy $\epsilon(\ell)$ are locally conserved and  hence satisfy  corresponding continuity equations. In the continuum limit,  these equations take the form 
\begin{align}
\f{\p r(x,t)}{\p t}&=\f{\p p(x,t)}{\p x}, \nn \\
\f{\p p(x,t)}{\p t}&=-\f{\p {P}(x,t) }{\p x},\nn \\
\f{\p e(x,t)}{\p t}&=-\f{\p}{\p x}[ p(x,t) {P}(x,t)]~, \label{eqm}
\end{align} 
where the label index $\ell$ has been denoted by the corresponding continuous variable $x$ and ${P}(x)=-V'(x)$ is the local force.
Assume that the system starts in a state of thermal equilibrium at zero total average momentum characterized by the temperature ($T=\beta^{-1}$) and pressure ($P$), which fix the 
the average energy and average stretch of the chain. 
The distribution corresponding to this ensemble is
\bea\label{measure} 
\mathcal{P}(\{p(x),r(x)\})=\prod_{x} \frac{e^{-\beta[p_x^2/2+V(r_x)+P r_x]}}{Z_x}~, 
\quad Z_x=\int_{-\infty}^\infty dp \int_{-\infty}^\infty dr e^{-\beta [ p^2/2+V(r)+P r]}~. 
\eea 

Since the fields $r(x,t),~p(x,t)$ and $e(x,t)$ satisfy continuity equations, they evolve slowly suggesting a slowly evolving local equilibrium picture. 
We consider small fluctuations of the conserved quantities about their equilibrium values,
 $u_1(x,t)=r(x,t)- \la r \ra_{eq}$, $u_2(x,t) = p(x,t)$ and  $u_3(x,t)=\epsilon(x,t) -\la \epsilon \ra_{eq}$. Inserting these into Eqs.~\eqref{eqm} one obtains 
$\p_t u_\alpha=-\p_x j_\alpha$, where $j_\alpha$ are the corresponding Euler currents which are functions of $u_\alpha$s. Expanding these currents to second order in     
the fields as $j_\alpha= \sum_\beta A_{\alpha \beta} u_\beta + \sum_{\beta,\gamma} H^\alpha_{\beta \gamma}  u_\beta u_\gamma$, 
and then adding dissipation and noise terms (to ensure 
thermal equilibration) one arrives at the following noisy hydrodynamic equations
\bea
\p_t u_\alpha= -\p_x  \left[ A_{\alpha \beta} u_\beta + H^\alpha_{\beta \gamma}  u_\beta u_\gamma - \p_x \widetilde{D}_{\alpha \beta} u_\beta + 
\widetilde{B}_{\alpha \beta} \xi_\beta \right]~,~~~~\label{EOM}
\eea
where repeated indices are summed over.
The noise and the dissipation matrices, $\widetilde{B},\widetilde{D}$, are related to each other by the
fluctuation-dissipation relation $\widetilde D C + C \widetilde{D} = \widetilde{B} \widetilde{B}^T$, 
where the matrix $C$ corresponds to equilibrium correlations and its elements are $C_{\alpha \beta}(x)= \la u_\alpha(x,0) u_\beta(0,0)\ra$. 

It is useful to define normal modes of the linearized equations (dropping $u^2$ terms in Eq.~\eqref{EOM}) through the 
transformation $(\phi_{-},\phi_0,\phi_+)=\vec{\phi} = R \vec{u}$, where the matrix $R$ acts only on the component index and diagonalizes $A$, \ie~ 
$R A R^{-1}={\rm diag}(-c,0,c)$. The diagonal form implies that there are two 
sound modes, $\phi_\pm$, traveling at speed $c$ in opposite directions and one stationary but decaying heat mode, $\phi_0$.
The quantities of interest are the equilibrium spatio-temporal 
correlation functions $C_{s s'}(x,t)=\la \phi_s(x,t) \phi_{s'}(0,0)\ra$, where $s,s'=-,0,+$. Because the modes separate linearly in time, one argues that at large times the off-diagonal matrix elements of the correlator are small compared to the diagonal ones and that the dynamics of the diagonal terms decouples into three single component equations. After including the non-linearity it is seen that  to leading order the equations for sound modes have self-coupling terms of the form $\phi_{\pm}^2$. These then have  the structure of the noisy Burgers equation, for which the exact scaling function, denoted by $f_\mathrm{KPZ}$, are known.  For the heat peak the self-coupling coefficient vanishes for any interaction potential. Thus one has to study the sub-leading corrections, and calculations using the mode-coupling approximation result in the symmetric L\'{e}vy walk distribution, with a cut-off at $x=ct$.  
While this is an approximation, it seems to be very accurate. For the generic case of non-zero pressure, \ie~$P\neq0$,
 which corresponds either to asymmetric inter-particle potentials or to an externally applied stress, the prediction
 for the left moving, resp. right moving, sound peaks and the heat mode are  
\begin{eqnarray} 
&&\hspace{-30pt} C_{--}(x,t)=   \f{1}{(\lambda_s t)^{2/3}}~ f_{\mathrm{KPZ}} \left[~ \f{(x + ct)}{(\lambda_s t)^{2/3}}~\right]~,\quad C_{++}(x,t)=   \f{1}{(\lambda_s t)^{2/3}}~ f_{\mathrm{KPZ}} \left[~ \f{(x - ct)}{(\lambda_s t)^{2/3}}~\right]~, \label{eqscalS}\\
&&\hspace{60pt} C_{00}(x,t) =  \f{1}{(\lambda_e t)^{3/5}} ~f^{5/3}_{\mathrm{LW}}\left[~ \f{x}{(\lambda_e t)^{3/5}}~\right]~,\label{eqscalE} 
\end{eqnarray}
where $f_{\mathrm{KPZ}}(x)$ is the KPZ scaling function discussed in \cite{Spohn14,Prahofer04}, and tabulated in \cite{Prahofertable}. The scaling function 
$f_{\mathrm{LW}}^\nu(x)$ is given by the Fourier transform of the L\'{e}vy characteristic function $e^{-|k|^\nu}$, with a cut-off at $x=ct$. The scaling parameters $\lambda_s$ and $\lambda_e$ are known explicitly. 
On the other hand for an even potential at zero pressure \emph{i.e.} $P = 0$, all self-coupling coefficients vanish. As a result the scaling solutions within mode-coupling approximation change and 
one obtains
\begin{eqnarray} 
  &&\hspace{-30pt}  C_{--}(x,t)= \frac{1}{{(\lambda^0_s t)}^{1/2}} f_{\mathrm{G}} \left[~ \f{(x + ct)}{(\lambda_s^0 t)^{1/2}}~\right]~,\quad C_{++}(x,t)= \frac{1}{{(\lambda^0_s t)}^{1/2}} f_{\mathrm{G}} \left[~ \f{(x - ct)}{(\lambda_s^0 t)^{1/2}}~\right]~,\label{eqscalSP0}\\
   &&\hspace{60pt} C_{00}(x,t) =  \f{1}{(\lambda_e^0 t)^{2/3}} ~f^{3/2}_{\mathrm{LW}}\left[~ \f{x}{(\lambda_e^0 t)^{2/3}}~\right]~,\label{eqscalEP0}
\end{eqnarray} 
 where $f_\mathrm{G}(x)$ is the unit Gaussian with zero mean. The scaling parameters $\lambda^0_s$ is not known from microscopics while $\lambda^0_e$ is known explicitly in terms of $\lambda^0_s$.

\begin{figure}
\includegraphics[width=0.75\linewidth]{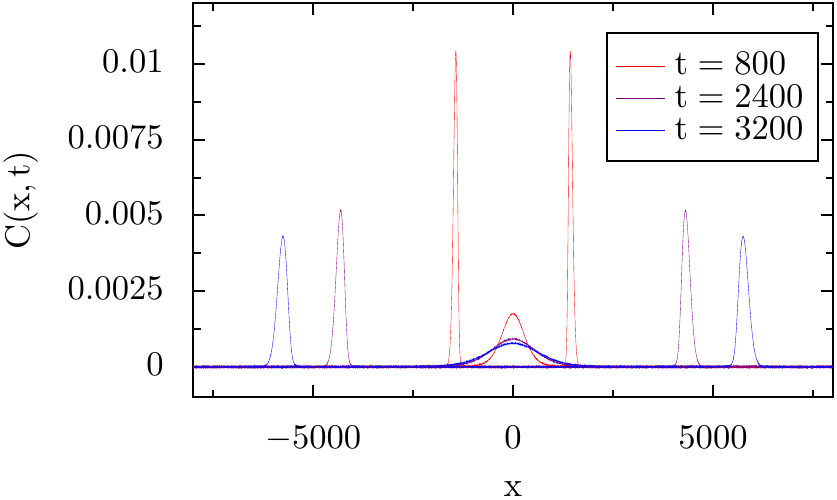}
\includegraphics[width=0.48\linewidth]{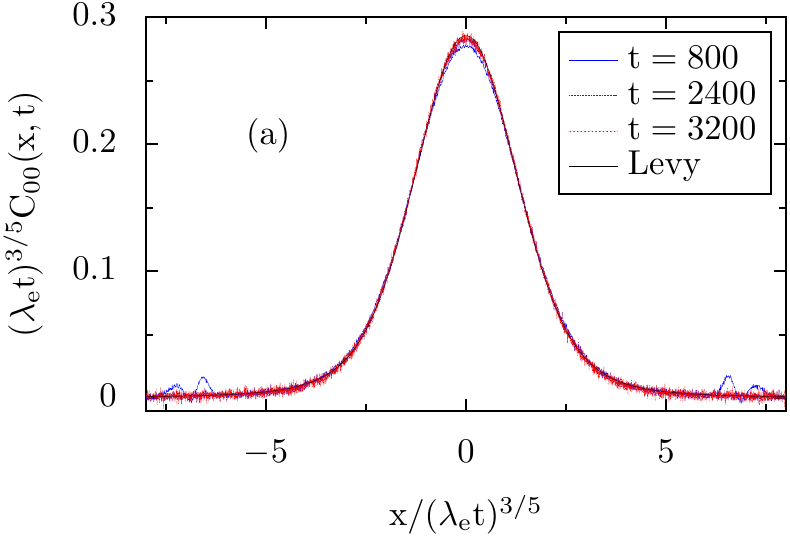}  \hskip 0.3cm
\includegraphics[width=0.48\linewidth]{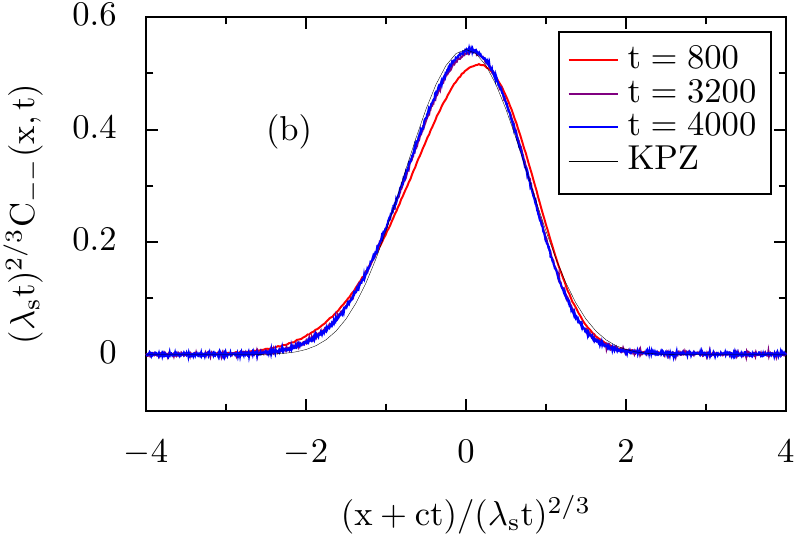}
\caption{ {\bf Top Panel}: Plots of the heat mode correlation  $C_{00}(x,t)$ (central peaks) and the sound mode correlations $C_{++}(x,t)$ and $C_{--}(x,t)$ (right and left moving peaks) in the FPUT chain, at three different times, for the parameter set with $k_2=1.0,k_3=2.0,k_4=1.0,T=5.0,P=1.0$  and system size $16384$. The speed of sound  was $c=1.80293$. We see that the heat and sound modes are well-separated. The numerical data in this plot were obtained by averaging over around $10^6-10^7$ initial conditions. 
{\bf Bottom Panel}: 
The heat mode (a) and the left moving sound mode (b) correlations, respectively, $C_{00}(x,t)$ and $C_{--}(x,t)$ are plotted 
at different times, using a L\'{e}vy-type-scaling for the heat mode and KPZ-type scaling for the sound mode. 
Here we observe a very good collapse of the data at different times. Moreover, 
we observe a good fit to the L\'{e}vy-stable distribution with $\lambda_e=5.86$ and a  reasonable fit to the KPZ scaling function, with $\lambda_s=0.46$. The parameters used in this plot are $k_2=1.0,k_3=2.0,k_4=1.0,T=5.0,P=1.0$. 	(Adapted from Das et.al. with permission from \cite{Das14} Copyright (2014) by American Physical Society).}
\label{S2scalA}
\end{figure}

Here we present molecular dynamics simulation results for the FPUT chain that were obtained in \cite{Das14} which verify the predictions of NFH. In Fig.~(\ref{S2scalA}, top panel), the two-point correlation functions $C_{00}(x,t),~C_{++}(x,t)$ and $C_{--}(x,t)$ are plotted as a function of $x$ for three values of time $t=800,~2400$ and $3200$. The parameters used in this plot are $k_2=1.0,k_3=2.0,k_4=1.0,T=5.0,P=1.0$ for which 
one gets $c=1.80293$ and we also see there is a good separation of the heat and sound modes.   In Fig.~(\ref{S2scalA}, bottom panel)  we also  find an excellent collapse of the heat mode and the sound mode data with the expected scalings . The scaled data for the heat mode  fits very well to the L\'{e}vy-scaling function whereas 
the same for the sound-mode still shows some asymmetry but is quite close to the KPZ function.  The numerically estimated values of the constants $\lambda_{s,e}$ are $\lambda_s=0.46$ and $\lambda_e=5.86$. These are in close agreement to the theoretically obtained values $\lambda_s=0.396$  and $\lambda_e=5.89$.

\section{Stochastic models: exact results on fractional equation description}
\label{sec:stochastic}

It  is now well understood that conservation laws  play an important role in 
observation of super-diffusive transport in one-dimensional systems. As we saw in the previous section, NFH provides some understanding of the emergence of  L\'{e}vy-walk behaviour, which seems to capture several aspects of anomalous transport. 
However, providing a completely rigorous microscopic derivation of the  L\'{e}vy-walk picture in a Hamiltonian model has been difficult, though there have been some attempts \cite{Mellet15}. While generic non-linear Hamiltonian models are difficult to analyze, analytical results have been obtained for harmonic chains whose Hamiltonian dynamics is perturbed by stochastic noise that breaks integrability of the system \cite{Basile06, Bernardin12, Saito18}.  These stochastic models attempt to mimic nonlinear chains and for these  models, several exact results both in the closed system set-up 
and the open system set-up have been obtained. In particular one can rigorously establish non-local response relation Eq.~\eqref{NLFL} and the fractional diffusion equation Eq.~\eqref{fraceq}.  There are two widely studied stochastic models which we discuss below.
  
  \subsection{Harmonic chain with volume exchange}
  \label{sec:HCVE}
 \noindent
This model is defined on a   one dimensional lattice  where each site carries a `stretch' variable $\eta_i,~i\in \mathbb{Z}$ and the energy of the system is  $E=\sum_i \eta_i^2$. The dynamics has two parts: (a) a deterministic part given by $\frac{d\eta_i}{dt}= \eta_{i+1}-\eta_{i-1}$ and (b) a stochastic exchange part where $\eta$s from any two randomly chosen neighboring sites,  are exchanged at a constant rate $\gamma$. We refer to this model as Harmonic chain with volume exchange (HCVE). This model was introduced in \cite{Bernardin12} where it was shown that the energy current auto-correlation decays  as $\sim 1/\sqrt{t}$,  implying super-diffusive transport. It is easy to see that this system has only two conserved quantities namely, the total ``volume" $\sum_i \eta_i$ and the total energy $\sum_i \eta_i^2$. The evolution of the density fields corresponding to these conserved quantities at the macroscopic length and time scales was studied in \cite{Spohn15} using NFH, where it has been shown that this model has two normal modes - one diffusive sound mode and a  $\frac{3}{2}$-asymmetric L\'evy heat mode. Subsequently, it was rigorously shown that the local energy density $e(x,t)$ satisfies a $(3/4)$-skew-fractional equation \cite{Bernardin16}
\begin{align}
\partial_t e(x,t)= -  \frac{1}{\sqrt{2\gamma}}~\mathbb{L}^v_\infty e(x,t),~~\text{for}~~x\in (-\infty, \infty),~~\text{where}~~\mathbb{L}^v_\infty=[(-\Delta)^{3/4}-\nabla (-\Delta)^{1/4}], \label{2comp-frac}
\end{align}
with $\Delta$ as the usual Laplacian operator. In the Fourier domain, defined  by $e(k,t) = \int_{-\infty}^\infty e(x,t) e^{\iu x k} dx$, the above equation reads as
 \eqa{
 	\partial_t e(k,t)= -  \frac{1}{\sqrt{2\gamma}}|k|^{3/2}[1-\iu~\text{sgn}(k)]~e(k,t).
 	\label{eta-frac}
 }
Note that for the diffusive case the analogue of the above equation would be $\partial_t e(k,t)= - D k^{2}~e(k,t)$. The above results suggest that,  in the open set-up where the system is connected to two reservoirs at different temperatures, this model would exhibit anomalous scaling of the steady state current $j$ with system size $N$. In  \cite{Bernardin12}, it has been numerically shown that indeed $j \sim 1/{\sqrt{N}}$. Recently, an understanding of the open system was achieved using the fractional equation description, which we now discuss \cite{Kundu18}. An aspect that we will point out here is that the fractional-equation-type description in the open-set up is strongly dependent on boundary conditions (fixed or free or mixed).

For the open system case, we  consider a finite lattice of size $N$,  connected to two thermal reservoirs at temperatures $T_\ell$ and $T_r$ on the left and right boundaries. The dynamics of the $\eta_i,~i=1,2,...,N$ now gets modified to 
\begin{align}
\begin{split}
\frac{d\eta_i}{dt}= ~\eta_{i+1}-\eta_{i-1} +\delta_{i,1}\left(-\lambda \eta_1 + \sqrt{2 \lambda T_\ell} \xi_\ell(t) \right) &+\delta_{i,N}\left(-\lambda \eta_N + \sqrt{2 \lambda T_r} \xi_r(t) \right)\\
&+\text{stochastic~exchange~at~rate~}\gamma.
\end{split}
\label{dynamics-eta}
\end{align}
The Langevin terms at the boundaries $i=1$ and $i=N$ appear due to the baths and $\xi_{\ell,r}(t)$ are  independent Gaussian white noises with mean zero and unit variance. We consider  fixed boundary conditions  $\eta_0=\eta_{N+1}=0$. 

Our main interest is to obtain an equation in this finite system, analogous to Eq.~\eqref{2comp-frac},  to describe the evolution equation of the temperature profile . To do this we first define the local temperature $T_i(t)=\langle \eta^2_i(t)\rangle$ and the off-diagonal correlations  $C_{i,j}(t)=\langle \eta_i(t)\eta_j(t)\rangle$, $i\neq j$,  which characterize the non-equilibrium state of the system. Interestingly, it turns out that  the equations for two point correlations do not depend on higher order correlations and this property leads to the model's solvability.  The evolution of these quantities in the bulk ($2 <i,j < N-1$) can be  obtained from Eq.~\ref{dynamics-eta} as:
  \begin{align}
  \begin{split}
  &\dot{C}_{ij}=C_{i+1,j}-C_{i-1,j}+C_{i,j+1}-C_{i,j-1}  +\gamma[C_{i-1,j}+C_{i+1,j}+C_{i,j-1}+C_{i,j+1}-4C_{i,j}], \\
  &\dot{C}_{i,i+1}=T_{i+1}-C_{i-1,i+1}+C_{i,i+2}-T_{i} +\gamma[C_{i-1,i+1}+C_{i,i+2}-2C_{i,i+1}], \\
  &\dot{T}_i=2[C_{i,i+1}-C_{i-1,i}]+\gamma[T_{i+1}+T_{i-1}-2T_i].
  \end{split}
  \label{corr-eq}
  \end{align}
The equations involving the boundary terms are given in \cite{Kundu18}. 
 Note that  in an infinite system, we get the same set of equations with $i,j \in \mathbb{Z}$.  For the finite open system, solving the above equations exactly seems to be difficult. However, it was observed numerically \cite{Kundu18}  that   for large $N$ the temperature field $T_i(t) $ scales as $T_i(t) = \mathcal{T} \left(\frac{i}{N},\frac{t}{N^{3/2}} \right)$ and  the correlation field $C_{i,j}(t)$ scales as 
  $
  C_{i,j}(t)= \frac{1}{\sqrt{N}}\mathcal{C}\left(\frac{|i-j|}{\sqrt{N}},\frac{i+j}{2N}, \frac{t}{N^{3/2}}\right), i \neq j.
  $
Inserting these into \eqref{corr-eq}, and expanding in powers of $1/\sqrt{N}$,   we find at leading order the following equations 
   \begin{align}
  \partial_v \mathcal{C}(u,v,\tau)&= - \gamma \partial_u^2  \mathcal{C}(u,v,\tau) 
  \label{evo-eq-1} \\
  \partial_v \mathcal{T}(v,\tau)&= - 2\gamma \left[ \partial_u  \mathcal{C}(u,v,\tau)\right]_{u=0} 
  \label{evo-eq-2} \\ 
  \partial_\tau \mathcal{T}(v,\tau) &= 2 \partial_v  \mathcal{C}(0,v,\tau),
  \label{evo-eq-3}
  \end{align}
where the scaling variables $u={|i-j|}/{\sqrt{N}},v={(i+j)}/{2N}, \tau = {t}/{N^{3/2}}$ are defined over $\{ 0\leq u \leq \infty~;~0 \leq v \leq~1~;~0 \leq \tau \leq \infty \}$. 

Note that for the isolated infinite system, one can follow the same procedure as above, but now replacing the scale parameter $1/N \to a$ where $a$ is the lattice spacing, to obtain the same set  of  Eqs.~\ref{evo-eq-1}-\ref{evo-eq-3} with a different domain $\{ -\infty \leq u \leq \infty~;~-\infty \leq v \leq~\infty\}$. These equations can be solved by Fourier transforms to get a skew fractional evolution equation for $\mathcal{T}(v,\tau)$  of the same form as Eq.~\eqref{eta-frac}. Defining Fourier transforms $\mc{T}(v,\tau)= \int_{-\infty}^\infty dk \hat{\mc{T}}(\tau) e^{-ik v}/(2 \pi)$ and $\mc{C}(u,v,\tau)= \int_{-\infty}^\infty dk \hat{\mc{C}}_k(u,\tau) e^{-ik v}/(2 \pi)$ in the variable $v$, we get
\begin{align}
\partial_u^2  \hat{\mathcal{C}}_k(u,\tau)&=\frac{i k}{\gamma}  \hat{\mathcal{C}}_k(u,\tau) \label{cqa} \\
i k \hat{\mc{T}}_k(\tau) &=  2\gamma \left[ \partial_u  \hat{\mathcal{C}}_k(u,\tau)\right]_{u=0} \label{cqb} \\
\partial_\tau \hat{\mathcal{T}}_k(\tau) &= -2 i k  \hat{\mathcal{C}}_k(0,\tau)~.
\label{cqc}
\end{align} 
Solving the first equation \eqref{cqa}, with the condition that correlations vanish at $u= \pm \infty$, we get
\eqa{
\hat{\mc{C}}_k(u,\tau)= A_k(\tau) \exp {\left[-(1+i \text{Sgn} (k)) \frac{|k|^{1/2}}{\sqrt{2 \gamma}} |u| \right] }  \label{cqd}
}
The equation \eqref{cqb} relates the constant $A_k$ to $\hat{\mc{T}}_k$:
\eqa{
A_k(\tau)=-\frac{ik(1-i \rm{Sgn} (k)) }{2\sqrt{2 \gamma |k|}} \hat{\mc{T}}_k.\label{cqe}
}    
Using Eqs.~(\ref{cqd},\ref{cqe}) in Eq.~\eqref{cqc} we get the infinite line equation in Eq.~(\ref{eta-frac}).  

We  now go back to  the open system case where the solution is more non-trivial. To solve these equations in the open set-up, we proceed as done for the regular diffusive heat equation, and write the solution as sum of a steady state part  and 
a relaxation part
  \begin{align}
  \mathcal{T}(v,\tau)&=\mathcal{T}_{ss}(v)+\mathcal{T}_r(z,\tau)  \label{T}\\
  \mathcal{C}(u,v,\tau)&=\mathcal{C}_{ss}(u,v)+\mathcal{C}_r(u,z,\tau), \label{C}
  \end{align}
where we have defined $z=1-v$. We note that under this transformation, the ``anti-diffusion''  Eq.~\eqref{evo-eq-1}, becomes a diffusion equation, with $v$ as the time variable and $z$ the space variable. The relaxation part satisfies the  equations given in Eqs.~(\ref{evo-eq-1},\ref{evo-eq-2},\ref{evo-eq-3}), while  the steady state part satisfies these equations but with $\p_\tau \mathcal{T}_{ss}(v)=0$. The boundary conditions for the steady state part are given by \cite{Kundu18}
  \begin{figure}
  	\centering
  	  	  	\includegraphics[width=0.48\linewidth]{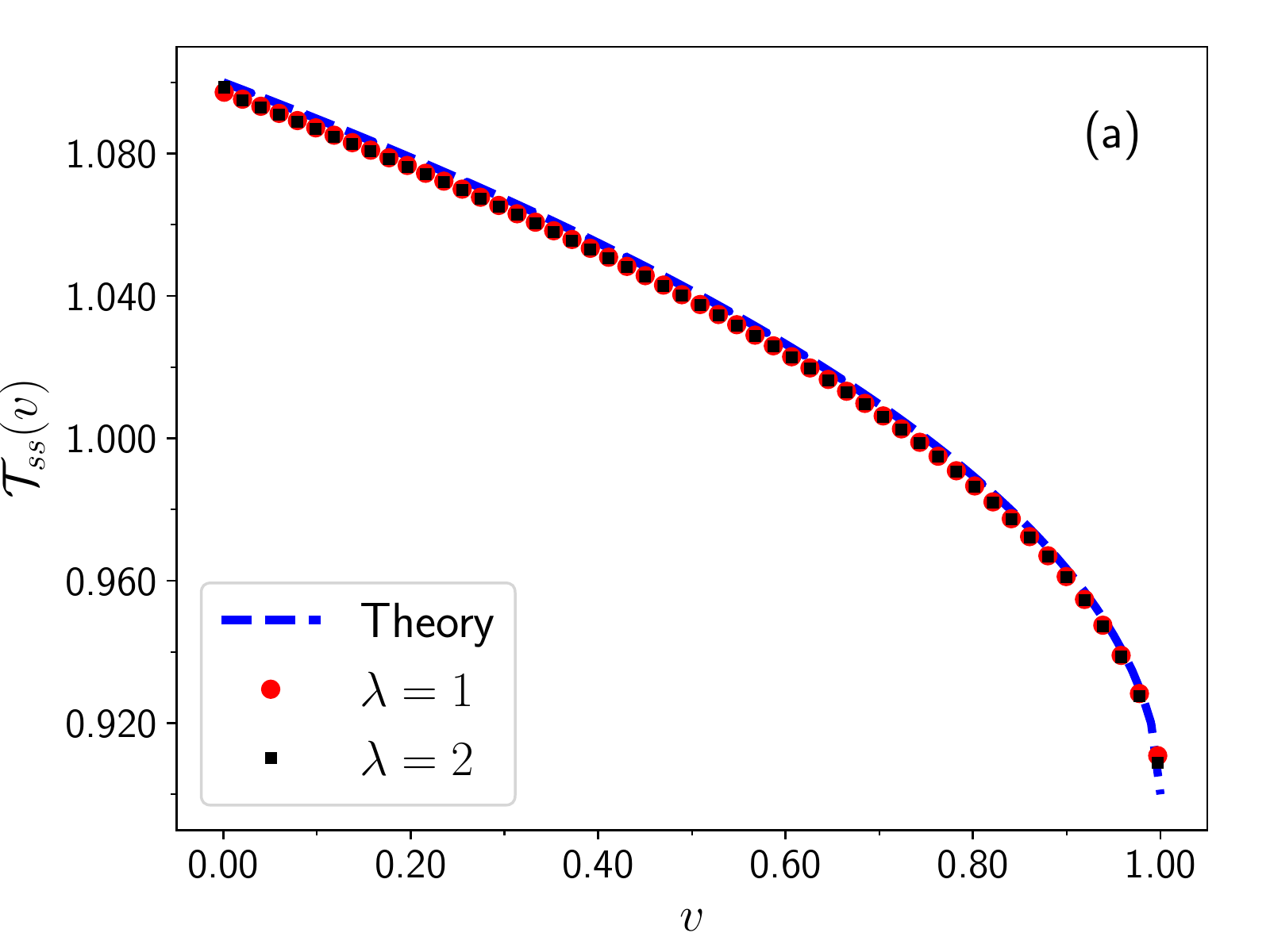}
  	\includegraphics[width=0.48\linewidth]{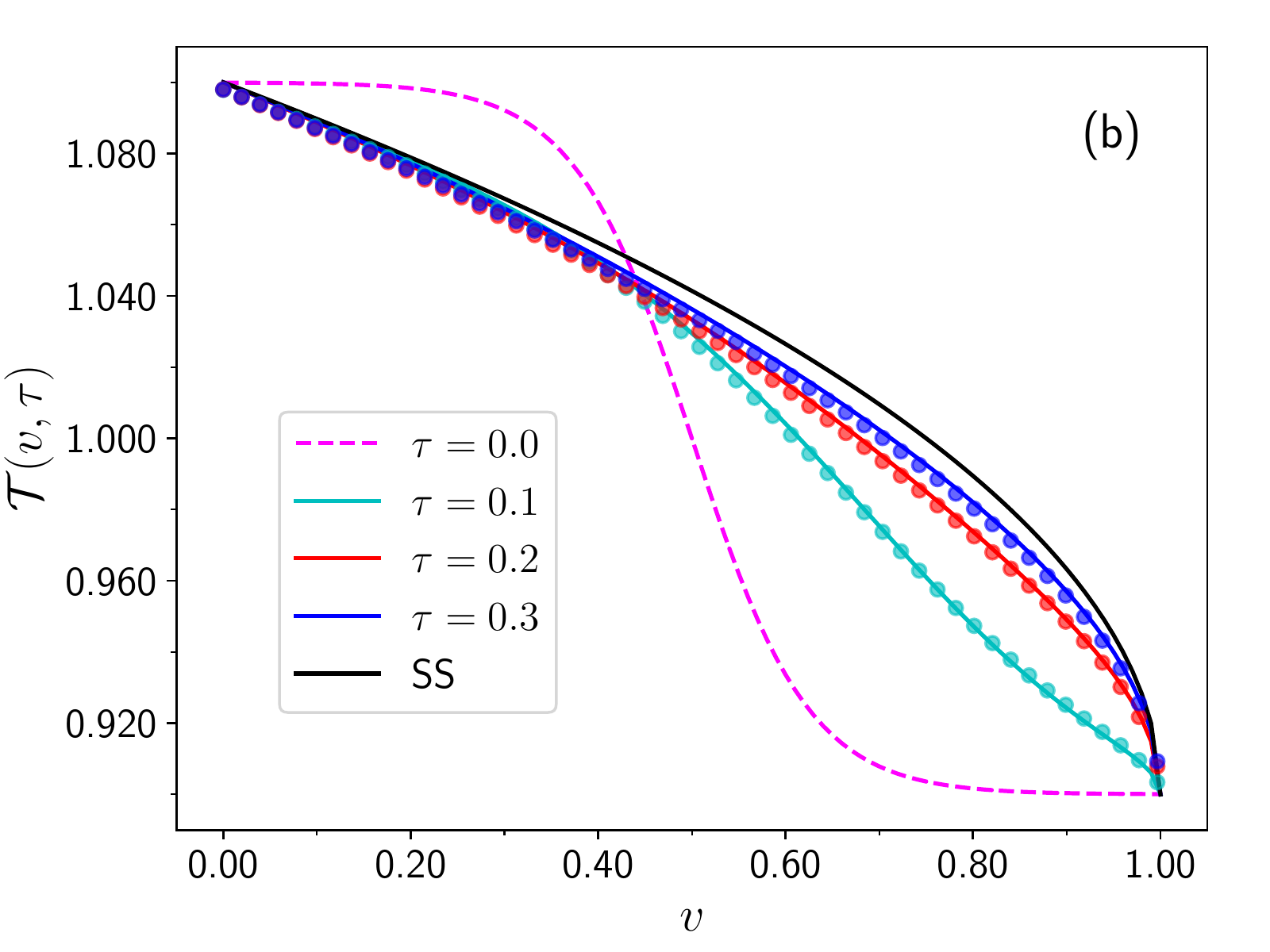}

  	\caption{(a) Steady state temperature of HCVE model ($\mc{T}_{ss}$) as given by Eq.\eqref{ss-soln} (blue dashed line) is compared with numerical simulations  for parameters $\omega = \gamma = 1, T_\ell = 1.1, T_r = 0.9$ and $N = 1024$, and for two different choice of $\lambda$. 
  		(b)  Numerical verification of the evolution of the temperature profiles $\mc{T} (v,\tau)$ given by Eq.~\eqref{hcve-Tr} (solid lines), starting from a non-staionary profile (dashed line). The points indicate simulation results for parameters are $\lambda= \gamma = 1, T_\ell = 1.1, T_r = 0.9$ and $ N = 2048.$ 	(Adapted from Kundu et.al. with permission from \cite{Kundu18} Copyright (2018) by American Physical Society).}
  	\label{fig:tevoa}
  \end{figure}
  \eqa{  ~\mc{C}_{ss} (u,z \to 0)&=0,~\mc{C}_{ss}(u \to \infty,z)=0,~\mc{C}_{ss}(u=0,z) = J/2. \nn \\
\mathcal{T}_{ss}(v=0)&=T_\ell,~ \mathcal{T}_{ss}(v=1)=T_r,~ 
  } 
where we have used Eq.~\eqref{evo-eq-3} to identify  $J= 2 \mc{C}_{ss}(u=0,z)$ as the  NESS current  which gets determined by the boundary conditions for $\mathcal{T}_{ss}(v)$. In terms of the original unscaled variables, the true  current is given by $j_{ss} = J/\sqrt{N}$. The solution of the steady state equations is given by \cite{Kundu18}
  \begin{align}
  \begin{split}
  \mathcal{T}_{ss}(v)&= T_r + (T_\ell-T_r)~\sqrt{1-v}, \\ 
  \mathcal{C}_{ss}(u,v)&= -\frac{T_\ell-T_r}{4} \sqrt{\frac{\pi}{\gamma}}~\text{erfc} \left( \frac{u}{\sqrt{4 \gamma(1-v)}}\right),\\
  J &= \frac{T_\ell-T_r}{2} \sqrt{\frac{\pi}{\gamma}}.
  \end{split}
  \label{ss-soln} 
  \end{align}	
  In Fig.~(\ref{fig:tevoa}a), we show a comparison of the above result for steady state temperature profile  with those obtained from direct simulations of the microscopic model, and we see very good agreement. It is interesting to note that the temperature profile is non-symmetric under space reversal as the microscopic model itself does not have such symmetry. This fact is also reflected in hydrodynamics where this shows in the existence of a single sound mode.

For the  relaxation part  we look for solutions which satisfy the initial condition $\mathcal{T}_r(z,0)$, $\mathcal{C}_r(u,z,\tau=0)=0$ and boundary conditions $ \mathcal{C}_r(u,z,\tau)|_{u \to \infty}=0$, $\mathcal{T}_r(0,\tau)=\mathcal{T}_r(1,\tau)=0$. The solution of the ``anti-diffusion''  Eq.~(\ref{evo-eq-1}), with $z$ as time variable,  with the boundary condition \eqref{evo-eq-2} can be obtained as \cite{Kundu18}
  \begin{flalign}
  &\mathcal{C}_r(u,z,\tau)=- \int_0^z \frac{\exp\left(-\frac{u^2}{4\gamma (z-z')} \right)}{\sqrt{4 \pi \gamma (z-z')}}~\frac{\partial \mathcal{T}_r(z',\tau)}{\partial z'}dz', \label{C-sol}.
  \end{flalign}
Using this in \eqref{evo-eq-2}  then gives finally the evolution equation for 
the temperature field 
  \begin{equation}
  \partial_\tau \mathcal{T}_r(z,\tau)= \frac{1}{\sqrt{\pi \gamma}} \partial_z \left[ \int_0^z dz' ~\frac{\partial_{z'}\mathcal{T}_r(z',\tau)}{\sqrt{z-z'}}~ \right],
  \label{heat-ano}
  \end{equation}
  inside the domain $0 \le z \leq 1$. 
   This is  a non-local equation which can be recognized as a continuity equation $\p_\tau \mathcal{T}_r(z,\tau)=-\p_z j(z,\tau)$ where the current $j$  is precisely in the form stated in Eq.~\eqref{NLFL}. This  is the open-system analogue of Eq.~\eqref{2comp-frac}. 
For the infinite system, a similar computation leads to  Eq.~\eqref{heat-ano} but with the lower limit of integration replaced by $z=-\infty$ and, by taking Fourier transforms, this can be shown to  reduce to Eq.~\eqref{eta-frac}.

\begin{figure}
	\centering
	\includegraphics[width=0.75\linewidth]{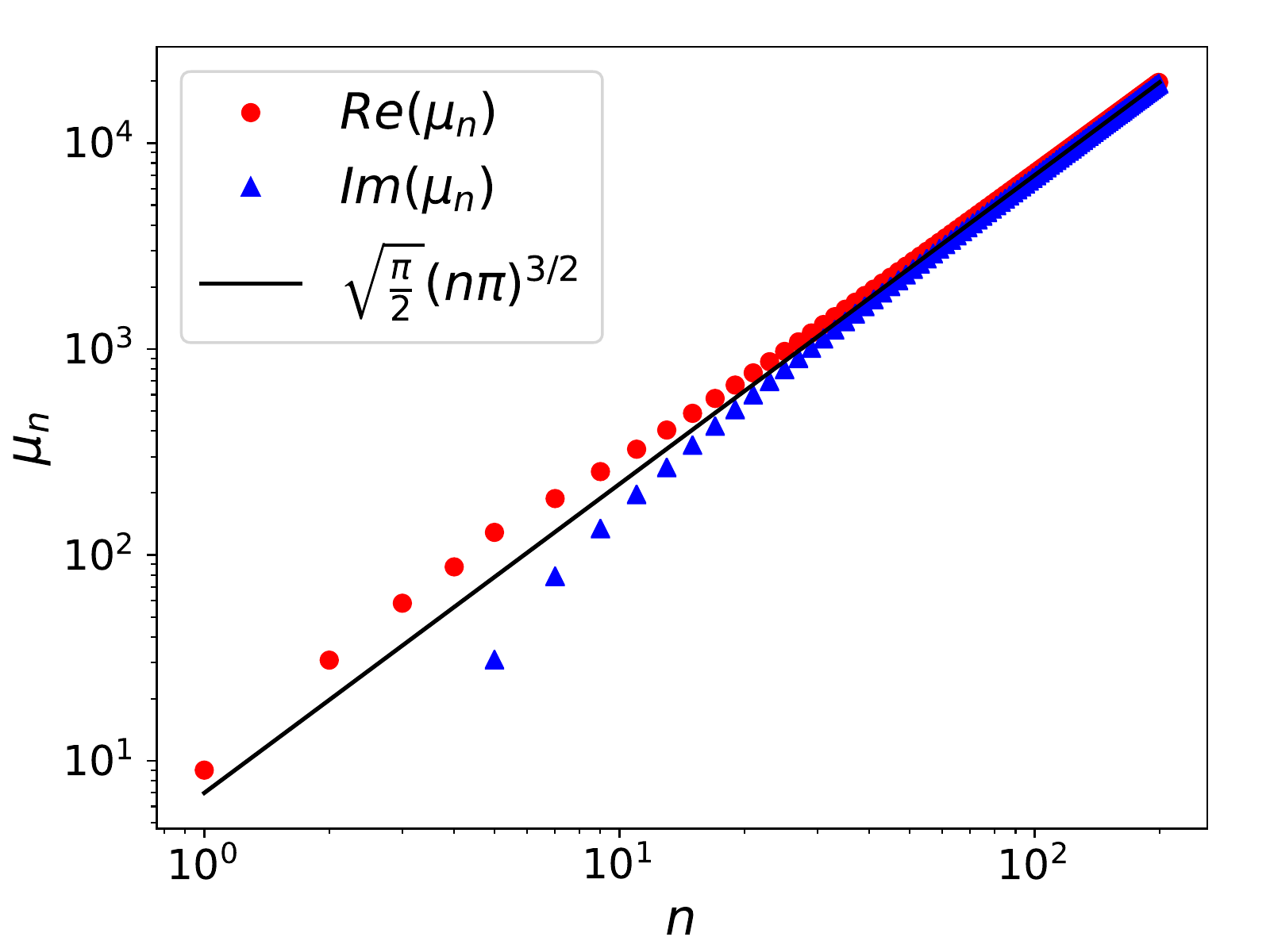}
	\caption{The real and imaginary part of the alternate eigenvalues for the matrix $\mathbb{L}^v$ in	
		Eq.~\eqref{eq:TempComp}. The first 4 eigenvalues are completely real and distinct. The higher eigenvalue comes in
		complex conjugate pairs. For large $\mu_n \sim (n\pi)^{3/2} 
		(1\pm i)$. For smaller $n$, there is a deviation from  asymptotic scaling due to finite definition of the operator. (Adapted from Kundu et.al. with permission from \cite{Kundu18} Copyright (2018) by American Physical Society).}
	\label{fig:eigenvalue}
\end{figure}

\begin{figure}
	\centering
	\includegraphics[width=1.0\linewidth]{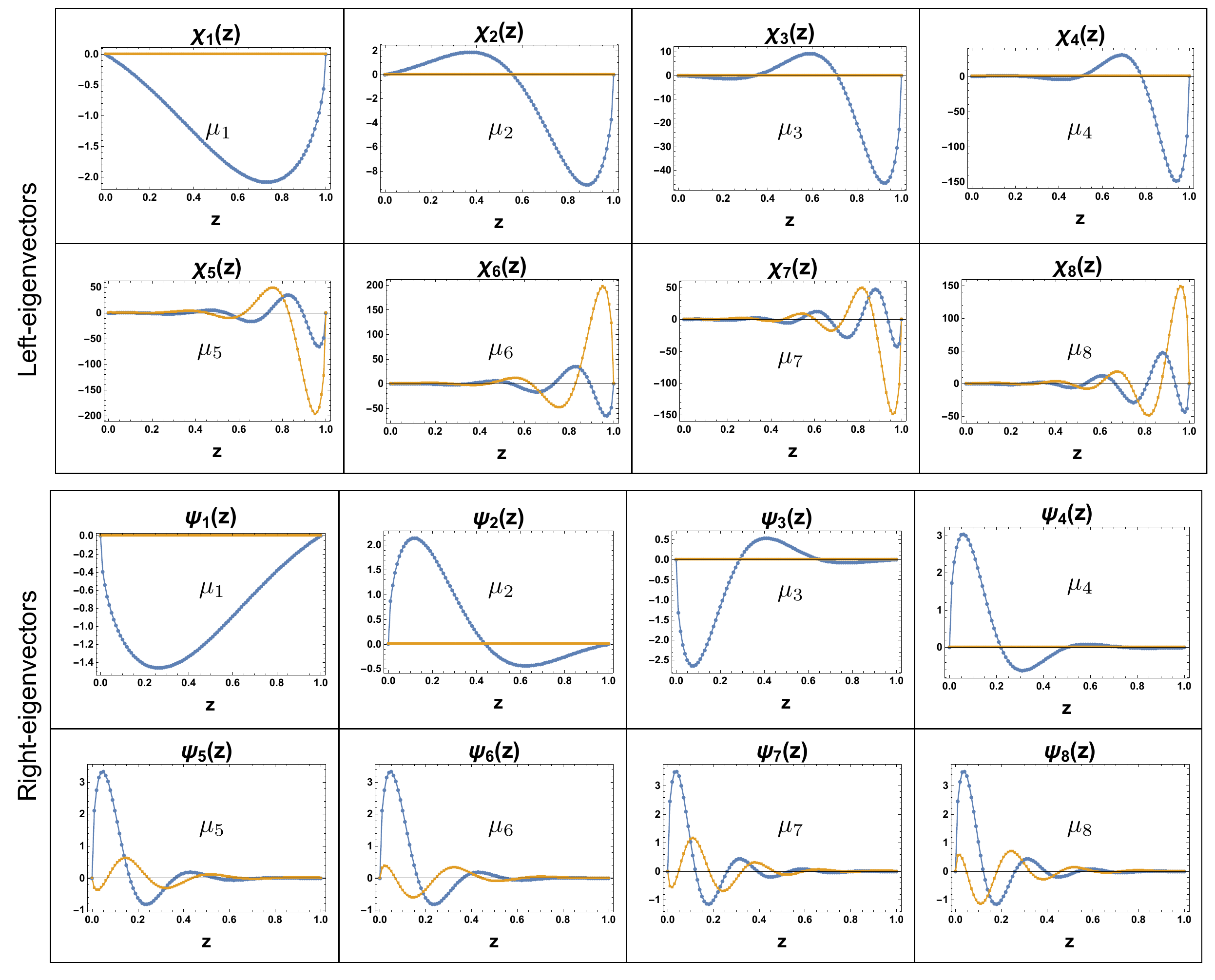}
	\caption{Left and right eigenvectors of matrix $\mathbb{L}^v$ in Eq.~\eqref{eq:TempComp} for $n=1,2,3,4,5,6,7,8$. The real parts are indicated by blue lines while orange denotes the imaginary part. Note that the eigenvectors corresponding to the real eigenvalues ($n=1,2,3,4$) are also real where the eigenvectors corresponding to complex eigenvalues ($n=5,6,7,8 ....$) are complex.
		(Adapted from Kundu et.al. with permission from \cite{Kundu18} Copyright (2018) by American Physical Society).}
	\label{fig:eigenvectors}
\end{figure}

We now proceed to solve Eq.~\eqref{heat-ano} to find the temperature evolution.
It is natural to expand the temperature profile $\mathcal{T}_r(z,\tau)$ in a basis set satisfying Dirichlet boundary conditions, and we choose the set $\alpha_n(z) = \sqrt{2} \sin (n \pi z),~n =1,2,3...$.   Substituting
$\mc{T}_r(z,\tau) = \sum_n \dx{\mathcal{T}}_n (\tau)\alpha_n(z)$ in  Eq.~\eqref{heat-ano}, we get
\eqa{ \sum_n \dot{\dx{\mathcal{T}}}_n \alpha_n(z) = \kappa~ \sum_n \dx{\mathcal{T}}_n (\tau) (n \pi) \p_z  \int_0^z \frac{ \phi_n(z')}{\sqrt{ z-z'}} dz'.}
Further we expand the function $f_n(z) = \p_z \int_0^z  { \phi_n(z')}/{\sqrt{ z-z'}} dz'  = \sum_{l=1} \zeta_{nl} \alpha_l(z)$ where 
$\zeta_{nl}=\int_0^1dz~f_n(z)~ \alpha_l(z)$.
Using orthogonality, we get
\eqa{\ket{\dot{\dx{\mathcal{T}}}} = \kappa~ \mathbb{L}^v\ket{\dx{\mathcal{T}}} , \label{eq:TempComp}} 
where $\mathbb{L}^v_{nl} = (n \pi)\zeta_{nl} $ and the column vector $\ket{\mathcal{T}}$ has elements $\dx{\mathcal{T}}_n = \iprod{\alpha_n}{\dx{\mathcal{T}}}$.
The above equation is an infinite-dimensional matrix representation of the non-local Eq.~\eqref{heat-ano}.
To solve this, we diagonalize the matrix  $\mathbb{L}^v$ as $\mb{R}^{-1}\mathbb{L}^v\mb{R} = {\bm \mu}$, which gives  the time dependent solution as $\ket{\dx{\mathcal{T}}(\tau)} = \mb{R} e^{\kappa {\bm \mu}  \tau}\mb{R}^{-1}  \ket{\dx{\mathcal{T}}(0)}$ where $\mb{R}_{n,l}=\la \alpha_n|\psi_l\ra $ denotes the $n$-th  element of the $l$-th right-eigenvector of the matrix $\mathbb{L}^v$ and the diagonal matrix $\mu$ contains the corresponding eigenvalue $\mu_l$. The matrix $\mathbb{L}^v$ is real but non-symmetric and it has left eigenvectors $\langle \chi_l|$  whose elements are given by   $\langle \chi_l| \alpha_n\ra =\mb{R}^{-1}_{l,n}$. The formal solution for the temperature field $\mathcal{T}_r(z,\tau)$ can then be written as
\eqa{ \label{hcve-Tr}
\mc{T}_r(z,\tau) = \sum_n \dx{\mathcal{T}}_n (\tau)\alpha_n(z) &=
\sum_{n,m,l} \mb{R}_{n,l} e^{\kappa \mu_l \tau} \mb{R}^{-1}_{l,m} \left[\int_0^1 dz' \mc{T}_r(z',0) \alpha_m(z')\right] \alpha_n(z)  \nn \\
&= \int_0^1 dz' \left[ \sum_l \psi_l(z) \chi_l(z')  e^{\kappa \mu_l \tau} \right] \mc{T}_r(z',0)~,
}  
where $\psi_l(z)=\iprod{ z}{\psi_l}= \sum_n \mb{R}_{nl} \alpha_n(z)$ and $\chi_l(z)=\iprod{\chi_l}{z} =\sum_n \mb{R}^{-1}_{ln} \alpha_n(z)$. Finding the eigenspectrum of the matrix $\mb{L}^v$ is a difficult problem as the matrix is infinite-dimensional and non-symmetric. However, one can truncate the matrix at some order and diagonalize it numerically, assuming that the spectrum converges with increasing truncation order. In \cite{Kundu18}  the authors used this apprach to compute the eigenspectrum and thereby study the time evolution of the temperature profile. This is shown in Fig.~(\ref{fig:tevoa}b). The spectrum  is shown in  Fig.~\eqref{fig:eigenvalue} where  it is seen that  for large $n$, $\mu_n \sim \sqrt{\frac{\pi}{2}}|n \pi|^{3/2}(1 \pm \iu )$ which is similar to the spectrum  of the non-local operator $\mathbb{L}^v$ in Eq.~\eqref{eta-frac} describing the evolution in infinite system. 
  In  Fig.~\eqref{fig:eigenvectors} we show the left and right eigenvectors $\chi_n(z) = \sum_{l=1} \mb{R}^{-1}_{nl} \alpha_l(z)$ and $\psi_n(z) = \sum_{l=1} \mb{R}_{ln} \alpha_l(z)$ respectively, corresponding to the first eight eigenvalues. 
One observes that the eigenvectors corresponding to the first four eigenvalues are  real whereas the eigenvectors corresponding to the eigenvalues with $n>4$ are complex and come in conjugate pairs.

  \subsection{Harmonic chain with momentum exchange}
  \label{sec:HCME}
  \noindent
In the previous section we  discussed transport in the HCVE model which has two conserved quantities, namely volume and energy. In this section, we discuss heat transport in the harmonic chain momentum exchange (HCME) model  which has three conserved quantities, namely  volume, momentum and energy, that are the same as the ones  in usual anharmonic chains with Hamiltonian dynamics \cite{Lepri16, Dhar08}. 
The model consists of a harmonic chain of  particles each of unit mass and described by the degrees of freedom   $q_i,p_i$, with $i \in \mathbb{Z}$, corresponding respectively to  position and momentum.  As for the HCVE system, the dynamics of the HCME model also has two parts: (i) the usual deterministic part given by the Hamiltonian equations $\dot{q}_i =~p_i~,~~\dot{p}_i = \omega^2(q_{i+1}-2q_{i} + q_{i-1}),  ~~ i \in \mathbb{Z}$, where $\omega$ is the strength of the harmonic interaction and (ii) a  stochastic part consisting of exchanges of momenta between  neighboring particles (chosen at random) occurring with rate $\gamma$.   In the absence of the stochastic exchange, the underlying Hamiltonian dynamics is integrable and the transport in this system is ballistic due to the absence of any scattering mechanism. The stochastic exchange introduces a momentum conserving scattering mechanism, which should make the transport behavior non-ballistic. However, it turns out that the stochastic mixing is not sufficient to make the transport behavior diffusive. It has been shown rigorously that the energy current correlation in equilibrium of an infinite chain decays as $t^{-1/2}$ similar to that in the HCVE model \cite{Basile06}. This, through the closed system GK formula in Eq.~\eqref{GKclosed}, implies the anomalous system size scaling of the steady state current as $j \sim N^{-1/2}$. 

The HCME dynamics conserves the following three quantities: (a) total stretch $\sum_i r_i$ where $r_i= q_{i+1} - q_i$, (b) total momentum $\sum_i p_i$ and (c) the total energy $\sum_i e_i$ with 
$e_i=p_i^2/2 + \omega^2 r_i^2/2$. As a consequence, the corresponding local densities evolve slowly in the macroscopic length and time scales. In \cite{Jara15}, it has been analytically shown that 
the local energy density $e(x,t)$ in the isolated system evolves according to the following fractional diffusion equation  
\begin{align} \label{frachcme}
\partial_t e(x,t) = -\bar{\kappa}(-\Delta)^{3/4} e(x,t),~~~~-\infty < x < \infty, ~~~
{\rm where}~\bar{\kappa}=\frac{\omega^{3/2}}{2 \sqrt{2 \gamma}}~,
\end{align}
and  the fractional operator in  Fourier  space is represented by $(-\Delta)^{3/4} e^{i k x}=|k|^{3/2}~e^{ik x}$. The NESS of this system was analyzed in detail in \cite{Lepri09,Lepri10,Delfini10} where the scaling $j \sim N^{-1/2}$ and a closed form for the nonlinear temperature profile were established. More recently the fractional-equation-type description of this system in the open set-up was further discussed in \cite{Kundu19}. We summarize below some of these results for the open system.   We first discuss the steady state and relaxation properties which is followed  by the discussion on the evolution of the fluctuations  and in the end we discuss the role of boundary conditions.

\subsubsection{Typical behaviour of temperature, current  and other correlations}
\label{sec:HCME-Av}
In the open system HCME set-up, the two ends are attached to two reservoirs at temperatures $T_\ell$ and $T_r$.  The equations of motion are now modified by adding Langevin forces to the $1$st and the $N$th  particles:
	\eqa{
	\dot{q}_i &= ~p_i~, \nn \\ 
	\dot{p}_i &= (1-\delta_{i,1}-\delta_{i,N}) \omega^2(q_{i+1}-2q_{i} +q_{i-1})
	+\delta_{i,1}[\omega^2(q_2-\zeta q_1) -\lambda p_1 + \sqrt{2\lambda T_\ell} \xi_1] \label{eom} \\
	&~~~+\delta_{i,N}[ \omega^2(q_{N-1}-\zeta q_{N}) -\lambda p_N + \sqrt{2\lambda T_r} \xi_N], 
	+\text{stochastic~exchange~of ~momenta~at~rate}~\gamma, \nn
} 
for $i=1,2\ldots,N$, where $\xi_{1,N}$ are  independent Gaussian white noises with mean zero and unit variance, $\lambda$ is the friction coefficient, and the parameter $\zeta$ has been introduced to describe different boundary conditions. Free boundary conditions correspond to $\zeta=1$  while fixed boundary conditions are given by $\zeta=2$. We will first discuss the fixed boundary case \emph{i.e.} $\zeta=2$. 

 We will be interested  not only in NESS properties such as the form of the temperature profile and the current scaling with system size but also in the temporal evolution of the  temperature from some arbitrary initial profile to the steady state form. As in the case of the HCVE model, the analytical tractability of the HCME system comes from the fact that the evolution of the two-point correlations is given by a closed set of equations. 
The two point correlations include ${\bf U}_{i,j}=\langle q_i q_j \rangle$,${\bf V}_{i,j}=\langle p_i p_j \rangle$, and ${\bf Z}_{i,j}=\langle q_i p_j \rangle$ and the  local temperature  defined as $T_i (t)=\la p_i^2\ra$ consists of the diagonal elements of ${\bf V}$. 
For these, one obtains a set of coupled linear equations, similar in form to Eqs.~\eqref{corr-eq}, which one needs to solve with appropriate boundary and initial conditions. The number of equations in this case is much larger than the HCVE case and hence it is even more difficult to solve them analytically for finite $N$.  Observations from  numerical solutions of these equations reveal \cite{Lepri10} that for large $N$,   the temperature field $T_i$ and   the  correlations ${\bf z}^+_{i,j}=\left({\bf Z}_{i,j}-{\bf Z}_{i-1,j}+{\bf Z}_{j,i}-{\bf Z}_{j-1,i} \right)/2$ show the following scaling behaviors: $T_i(t) = \mc{T}(i/N,t/N^{3/2})$ and ${\bf z}^+_{i,j} = \frac{1}{\sqrt{N}} \mc{C}\left((|i-j|)/{N}^{1/2},(i+j)/2N,t/N^{3/2}\right)$. Hence, for large $N$ it is instructive to construct solutions of these scaling forms. Inserting these scaling forms in the discrete equations  of the two point correlations and taking the large $N$ limit one finds, at leading order in $1/\sqrt{N}$, the following partial differential equations  \cite{Lepri10}
\eqa{
	\gamma^2 \p_u^4 \mc{C}(u,v,\tau) &= \omega^2 \p_v^2 \mc{C}(u,v,\tau), \label{PDE-hcmea} \\
	\p_v T(v,\tau) &= -2 \gamma \p_u \mc{C}(u,v,\tau)\big|_{u\to0}, \label{PDE-hcmeb}\\
	\p_\tau T(v,\tau) &=  \omega^2 \p_v \mc{C}(u,v,\tau)|_{u\to0}, \label{PDE-hcmec}
}
where the scaling variables $u={|i-j|}/{\sqrt{N}},v={(i+j)}/{2N}, \tau = {t}/{N^{3/2}}$ are defined over the domain $u \in [0,\infty)$ and $v \in [0,1]$ with boundary conditions $\mc{C}(u,0,\tau)=\mc{C}(u,1,\tau)=0,\mc{C}(\infty,v,\tau)=0, \partial_u^3C(0,v,\tau)=0$ and $T(0,\tau)=T_\ell$ and $T(1,\tau)=T_r$.
We again note that for the isolated infinite system, one can follow the same procedure as above, but now replacing the scale parameter $1/N \to a$ where $a$ is the lattice spacing, to obtain the same set  of  Eqs.~\ref{PDE-hcmea}-\ref{PDE-hcmec} with a different domain $\{ -\infty \leq u \leq \infty~;~-\infty \leq v \leq~\infty\}$. Defining Fourier transforms $\mc{T}(v,\tau)= \int_{-\infty}^\infty dk \hat{\mc{T}}_k(\tau) e^{-ik v}/(2 \pi)$ and $\mc{C}(u,\tau)= \int_{-\infty}^\infty dk \hat{\mc{C}}_k(u,\tau) e^{-ik v}/(2 \pi)$ in the variable $v$, we get
\begin{align}
\partial_u^4  \hat{\mathcal{C}}_k(u,\tau)&=-\frac{\omega^2k^2}{\gamma^2}  \hat{\mathcal{C}}_k(u,\tau), \label{cqpa} \\
i k \hat{\mc{T}}_k(\tau) &=  2\gamma  \partial_u  \hat{\mathcal{C}}_k(u,\tau)\big|_{u=0},\label{cqpb} \\
\partial_\tau \hat{\mathcal{T}}_k(\tau) &= - i k \omega^2 \hat{\mathcal{C}}_k(0,\tau)~.
\label{cqpc}
\end{align} 
Solving the first equation \eqref{cqpa}, with the condition that correlations vanish at $u =\pm \infty$, $ \p^3_{u}\mc{C}(u=0,v,\tau)=0$ and requiring that $\hat{\mathcal{C}}_k(u,\tau)$ is real [since $\mathcal{C}(u,v,\tau)=\mathcal{C}(u,-v,\tau)$], we get
\eqa{
\hat{\mc{C}}_k(u,\tau)= A_k(\tau) \left[\cos \left(\sqrt{\frac{\omega}{2 \gamma}} |k|^{1/2} u\right) - \sin \left(\sqrt{\frac{\omega}{2 \gamma}} |k|^{1/2} u\right) \right]\exp \left({-\sqrt{\frac{\omega}{2 \gamma}} |k|^{1/2} |u|}\right)  \label{cqpd}
}
The equation \eqref{cqpb} relates the constant $A_k$ to $\hat{\mc{T}}_k$:
\eqa{
A_k(\tau)=-\frac{ik}{2 \sqrt{2 \gamma \omega |k|}} \hat{\mc{T}}_k.\label{cqpe}
}    
Using Eqs.~(\ref{cqpd},\ref{cqpe}) in Eq.~\eqref{cqpc} we get 
\eqa{
\partial_\tau \hat{\mathcal{T}}_k(\tau)= -\frac{\omega^{3/2}}{2\sqrt{2 \gamma}}|k|^{3/2}  \hat{\mathcal{T}}_k(\tau)~,
}
which is the Fourier representation of Eq.~(\ref{frachcme}), with $\bar{\kappa}={\omega^{3/2}}/{2 \sqrt{2 \gamma}}$.

We  now go back to  the open system case where the solution is more non-trivial.
The boundary conditions for this case are given by $\mc{C}(u,0,\tau)=\mc{C}(u,1,\tau)=0,\mc{C}(\infty,v,\tau)=0, \partial_u^3C(0,v,\tau)=0$ and $T(0,\tau)=T_\ell$ and $T(1,\tau)=T_r$ (see \cite{Kundu19, Lepri10}). Note that the  domain of the $v$ variable in \cite{Lepri10} is $v \in (-1,1)$.  

\begin{figure}[t]
	\centering
	\includegraphics[width=\linewidth]{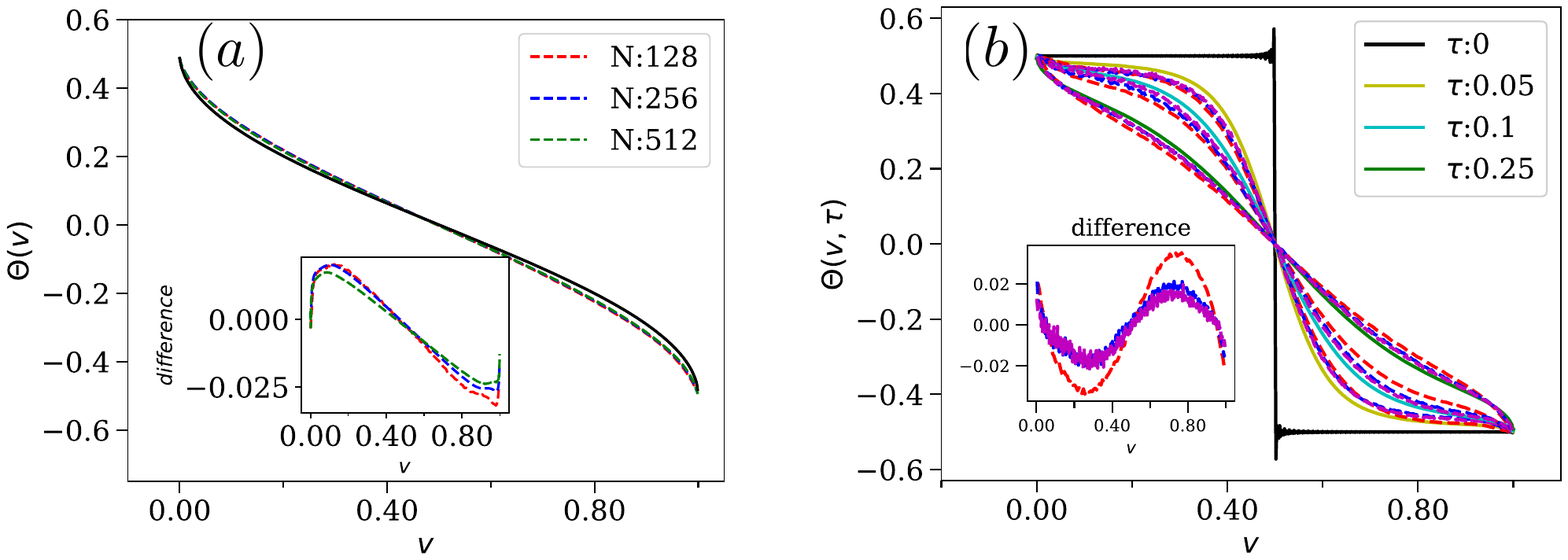}
	\caption{ (a) Comparison of temperature profiles obtained theoretically from Eq.~\eqref{THCME-SS} (solid black line) with the same obtained from direct numerical simulations of microscopic system for $N = 128, 256, 512$. The agreement between theory and numerics becomes better for larger $N$ as can be seen in the inset where the difference between theoretical curve (Eq.~\eqref{THCME-SS} and simulation data are plotted for various system sizes. (b) Time evolution of temperature starting from an initial step profile. The function $\Theta(v,\tau) = [\mc{T}(v,\tau)-\bar{T} ]/\Delta T$, with $\mc{T}(v,\tau)$ given by Eq.~\eqref{mcTevol}, is plotted and compared with direct numerical simulations. The dashed lines indicate simulation results for the 	time-evolution in HCME at different scaled times ($\tau$), for system sizes $N = 128$ (red), $N = 256$ (blue), $N = 512$ (magenta), while the solid lines are obtained from the theory.  The boundary temperatures were fixed at $T_\ell =1.5$ and $T_r = 0.5$. (Adapted from  Kundu et.al. with permission from \cite{Kundu19} Copyright SISSA Medialab Srl, IOP Publishing).}
	\label{fig:HCMEtemp}
\end{figure}

In the steady state, the analytical solutions of these equations [with $\partial_\tau T(v,\tau)=0$] were  obtained in \cite{Lepri09} and are given by 
\eqa{
	\mc{T}_{ss}(v) &= \overline{T} + \Delta T \Theta(v),~~\text{where}~~\Theta(v)=\frac{ \pi^{3/2}}{[\sqrt{8}-1  ] \zeta(3/2)} \sum_{n \; odd} \frac{\phi_n(v)}{\lambda_n^{3/4}}, \label{THCME-SS} \\
	\mc{C}_{ss}(u,v)&=-\frac{\Delta T \sqrt{\pi}}{2 \sqrt{\omega \gamma}  [\sqrt{8}-1  ] \zeta(3/2)} \sum_{n \; odd}  \exp \left(-\sqrt{\frac{n \pi \omega}{2 \gamma}} u\right) 
\left[ \cos \left(\sqrt{\frac{n \pi \omega}{2 \gamma}} u\right)-\sin \left(\sqrt{\frac{n \pi \omega}{2 \gamma}} u\right) \right] \frac{\sin(n \pi v)}{n},
	\label{eq:steadystate}}
where $\overline{T} = {(T_\ell+T_r)}/{2}$,  $\Delta T = T_\ell -T_r$ and $\phi_n(v) =\delta_{n,0}+(1-\delta_{n,0}) \sqrt{2}\cos(n \pi v)$ for $n \ge 0$. 
From Eq.~(\ref{PDE-hcmec}) we see that the current $J=-\omega^2 \mc{C}(u,v,\tau)|_{u\to0}$ is given by
\eqa{\label{JHCME-SS}
J=\frac{ (\omega \pi)^{3/2}}{8 \sqrt{\gamma}  [\sqrt{8}-1  ] \zeta(3/2)}\Delta T~.
}
Note that both the temperature profile and the current are independent of the friction coefficient $\lambda$. This is true only for the special case of fixed boundary conditions. 
Note also that the temperature profile in the steady state is intrinsically non-linear as can be seen in Fig.~(\ref{fig:HCMEtemp}a) where one   observes excellent agreement with data from simulations of the microscopic dynamics in Eq.~(\ref{eom}). It can be shown that  the temperature profile at both boundaries scales as $\sim (\delta v)^{\mu}$ with $\mu =1/2$ where $\delta v$ is the distance from the boundary \cite{Lepri09}. This singular behavior of $T_{ss}(v)$ is a common signature of anomalous transport and it is characterized by the meniscus exponent $\mu$. The value of $\mu$ however is non-universal and  depends strongly on the boundary conditions. We will discuss this in Sec.~\eqref{sec:BC}.  

To solve for the approach toward the above steady state results, we proceed as for the HCVE model. Separating the relaxation part and the steady state part we write
\eqa{
	\mc{C}(u,v,\tau) &= \mc{C}_{ss}(u,v)+ \mc{C}_r(u,v,\tau),  \\
	\mc{T}(v,\tau) &= \mc{T}_{ss}(v)+ \mc{T}_r(v,\tau).
}
Since the relaxation parts satisfy Dirichlet boundary conditions $\mc{C}_r(u,0,\tau)=\mc{C}_r(u,1,\tau)=0$ and $ \mc{T}_r(0,\tau)=\mc{T}_r(1,\tau)=0$, we expand them in the Dirichlet basis $\alpha_n(v)=\sqrt{2}\sin (n \pi v)$ for $n=1,2,3,...$ as
\eqa{
 \mc{C}_r(u,v,\tau)&=\sum_{n=1}^\infty \hat{\mc{C}}_n(u,\tau) \alpha_n(v), \\
 \mc{T}_r(v,\tau)&= \sum_{n=1}^\infty \hat{\mc{T}}_n(\tau) \alpha_n(v)~. \label{CTexp2} 
}
After inserting these expansions in Eqs.~(\ref{PDE-hcmea}-\ref{PDE-hcmec}) and using the orthogonality property of the $\alpha_n(v)$ functions, one gets the following (infinite order) matrix equation for the evolution of the components $\hat{\mc{T}}_n$: 
	\eqa{
	\dot{\hat{\mc{T}}}_n =-\bar{\kappa}	\sum_{l=1}^\infty  \flapfinitem_{nl} \hat{\mc{T}}_l,~~~~n=1,2,\ldots,\infty,~ 
	~\text{where: }~ \flapfinitem_{nl}=  \left[\mc{S} \Lambda^{3/4} \mc{S^\dagger} \right]_{nl}, \label{Lnk}
}
with $\mc{S}_{nl}=\iprod{\alpha_n}{\phi_l}=\int_0^1 dz \alpha_n(z) \phi_l(z)$,   $\Lambda_{n l} = \lambda_{n}\delta_{n l}$  is a diagonal matrix with $\lambda_n=(n\pi)^2$ and  the constant $\bar{\kappa} = {\omega^{3/2}}/{(2\sqrt{2 \gamma})}$. In the position basis, the above equation can be written as   
\eqa{\label{hcmeTev}
\p_\tau \mc{T}(v,\tau) = -\bar{\kappa} \flapfinitem \mc{T}(v,\tau),
}
 where the operator $\mb L^{p}$ is represented as 
\[\flapfinitem_{nl}=\left[\mc{S} \Lambda^{3/4} \mc{S^\dagger} \right]_{nl} =\bra{\alpha_n}\left[\sum_{m=0}^\infty \lambda_m^{3/4} \ket{\phi_m}\bra{\phi_m}\right] \ket{\alpha_l},~~\forall~~n,l= 1,2,\ldots, \infty\]
From this representation  one can identify the action of  $\flapfinitem$ on the set of basis functions $\phi_m$ (which satisfy Neumann boundary conditions) \cite{Lepri16,Kundu19}
\eqa{\flapfinitem \ket{\phi_m}= \lambda_m^{3/4} \ket{\phi_m}~. \label{L-in-phi}}

\begin{figure}
	\centering
	\includegraphics[width=0.49\linewidth]{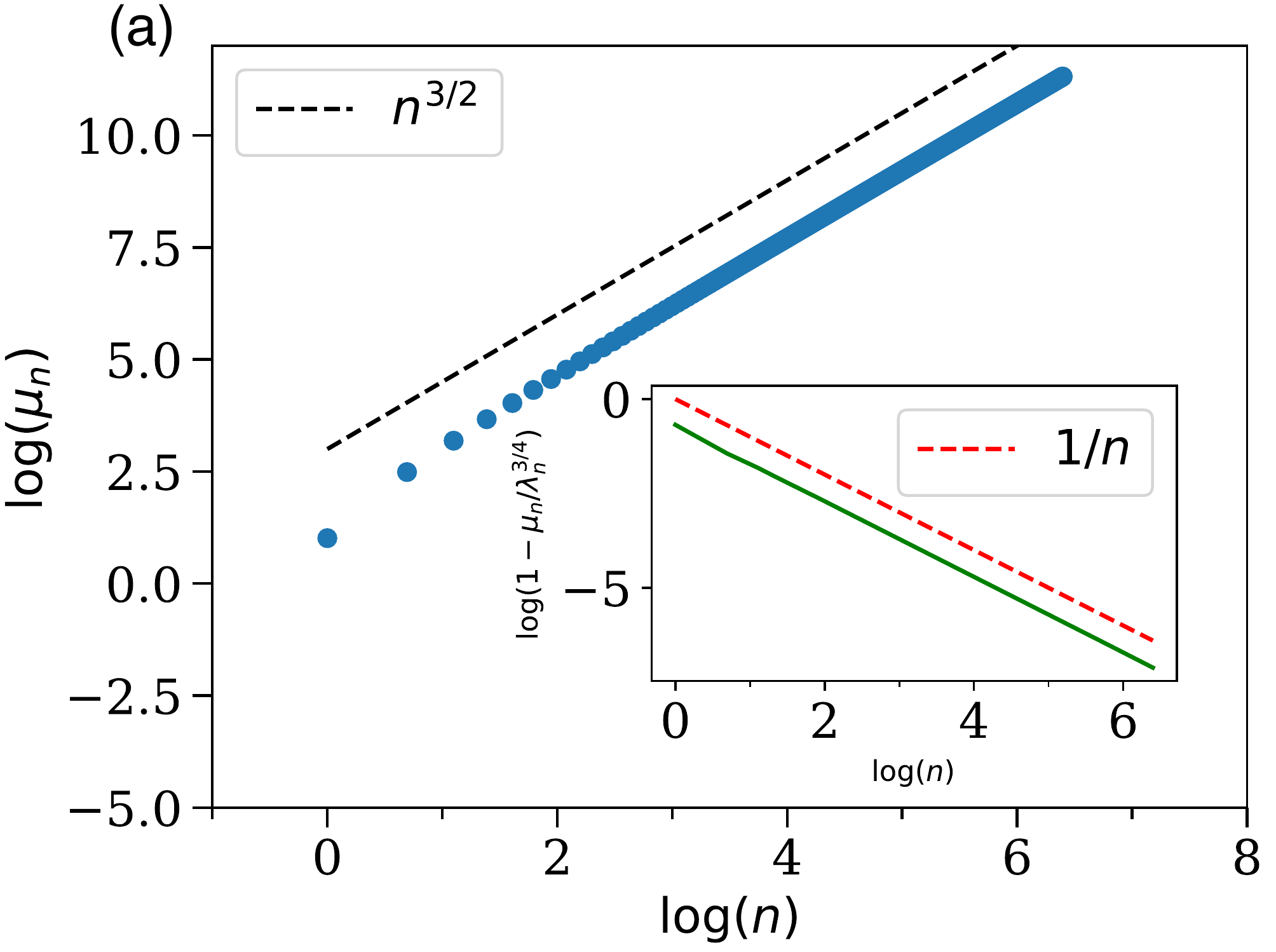}
	\includegraphics[width=0.49\linewidth]{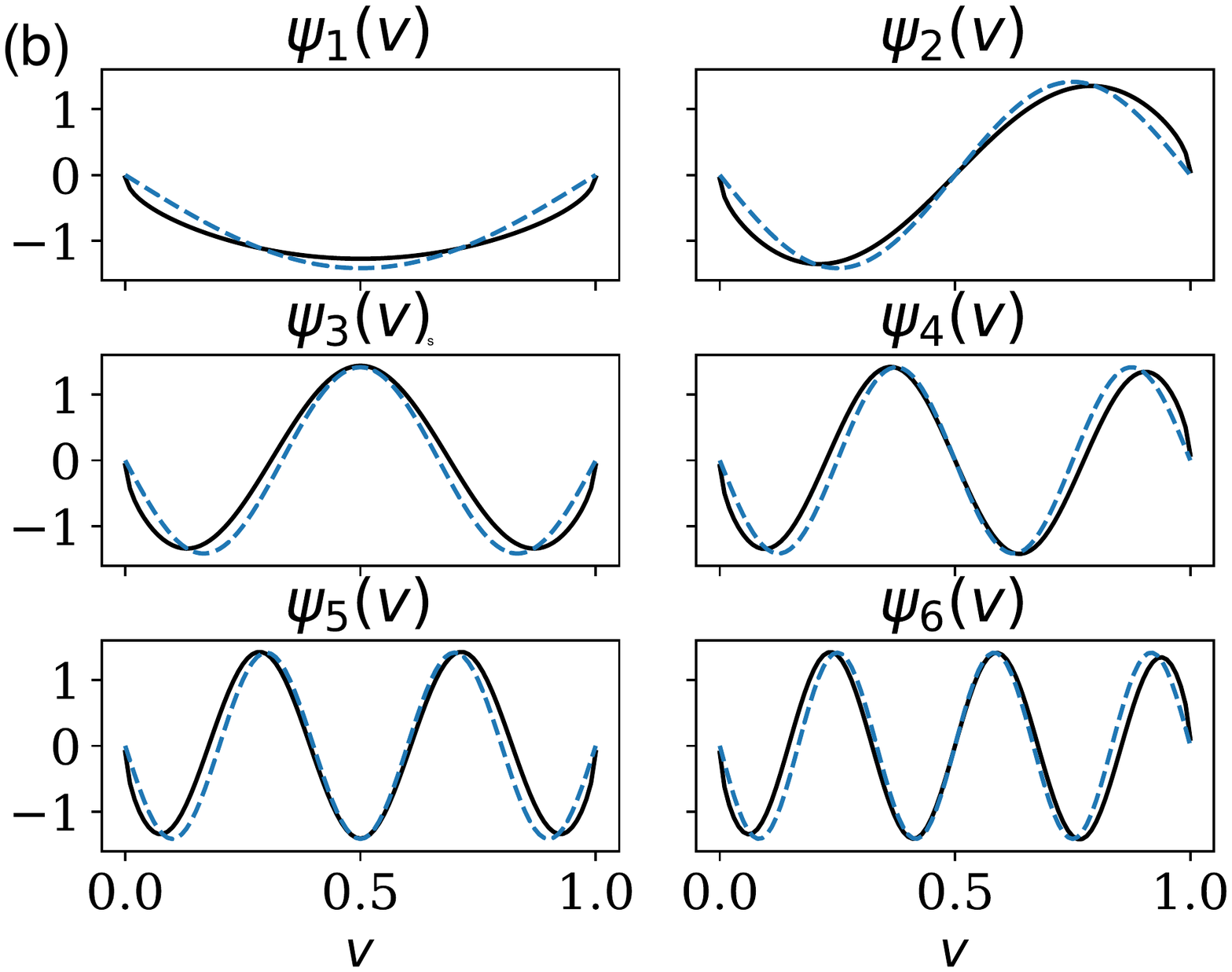}
	\caption{(a)Eigenvalues of the fractional operator in Eq.~\eqref{Lnk} corresponding to Dirichlet boundary conditions. For large $n$, the slope is seen to approach that of $n^{3/2}$ (black dashed line).  For small
		$n$ there is a systematic difference between the Dirichlet and Neumann eigenvalue snd   the inset plots the difference between the two. For large $n$ the difference between the two decreases inversely with $n$. (b) The first six eigenvectors, $\psi_n(v)$ (black lines), are compared to the corresponding Dirichlet eigenfunctions of Laplacian i.e. $\sin$-functions (blue dashed lines). The eigenstates are different from $\sin$-functions, especially near the boundaries, even for large $n$. (Adapted from  Kundu et.al. with permission from \cite{Kundu19} Copyright SISSA Medialab Srl, IOP Publishing.)}
	\label{fig:HCMEeigen}
\end{figure}

For the time evolution we need the eigenspectrum of $\flapfinitem$ with Dirichlet boundary conditions. The eigenstates $\psi_n(y)$ and eigenvalues $\mu_n$ can be obtained by diagonalizing the matrix $\mb{L}^p_{nm}$  in Eq.\eqref{Lnk}.  In \cite{Lepri10} the spectrum was obtained numerically by diagonalizing truncated form of the infinite-dimensional matrix $\mb{L}^p$. An alternate method was recently proposed in \cite{Kundu19} which gives the spectrum directly as roots of a transcendental equation and explicit series form expressions for the wave functions in the $\phi_n$ basis. 
The numerical values of the computed eigenvalues are plotted in Fig.~(\ref{fig:HCMEeigen}a), where we see that for large $n$ the eigenvalues scale as $\mu_n \approx (n \pi)^{3/2}$.  At smaller values $n$ there is a systematic deviation from the Neumann spectrum, $\lambda_n$, for example the first three eigenvalues ($\mu_n$) are given by  $\mu_1 \approx 2.75,\mu_2 \approx 12.02, \mu_3 \approx 24.22$. 
As  shown in the inset of Fig.~(\ref{fig:HCMEeigen}a) the relative difference between $\mu_n$ and $\lambda_n$ decreases as $1/n$.   
 The first few numerically computed eigenvectors are shown in Fig.~(\ref{fig:HCMEeigen}b) where they are compared with the basis functions $\alpha_n$ which are the Dirichlet eigenfunctions of the usual Laplacian.  We observe that they are different  and in particular show a non-analytic behavior at the boundaries.  For example  near the  boundaries one finds $\psi_n(\delta v) \sim \sqrt{\delta v}$,  where $\delta v$ is the distance from the boundaries. 
 The eigenspectrum of fractional operator in a bounded domain, with different boundary conditions, has  been 
 discussed earlier in the literature, using somewhat heuristic approaches   \cite{Zoia07,Buldyrev01,Buldyrev01a,Chen04}. However their connection  to the spectrum of $\mathbb{L}^p$ defined here is unclear.

Using these Dirichlet eigenvalues and eigenfunctions, we follow the steps leading to Eq.~(\ref{hcve-Tr}) and  obtain the  following for the time evolution of an arbitrary initial profile:
\eqa{\label{mcTevol}
	\mc{T}_r(v,\tau) =  \int_0^1 dv' \left[ \sum_{l=1}^\infty \psi_l(v) \psi_l(v')  e^{-\bar{\kappa} \mu_l \tau} \right] \mc{T}(v',0)~.
}  
In Fig.~(\ref{fig:HCMEtemp}b), a numerical verification of the above time evolution is shown. We note that Eq.~\eqref{hcmeTev} can be cast in the form of a continuity equation $\partial_\tau \mc{T}_r(v,\tau)=-\p_v j(v,\tau)$ with $j$ in the form \cite{Kundu19}
\eqa{
j(v,\tau)= -\bar{\kappa} \int_0^1 d v' \mc{K} (v,v') \p_{v'} \mc{T}(v',\tau)~, \label{j-nonL-hcme}
 }  
where the kernel $\mc{K}$ is defined through it's action on a test function $g(v)= \sum_{n=1}^\infty g_n \alpha_n(v)$ 
\eqa{
\label{defKhcmeFi}
\int_0^1 dv' \mc{K}(v,v') g(v')= \sum_{n=1}^\infty \frac{1}{\sqrt{n \pi}} g_n \alpha_n(v)~.
} 
The operator $\mb{L}^p$ can be expressed in terms of $\mc{K}$ as
\eqa{
\label{mbL-K-rel}
\la v | \mb{L}^p |v'\ra = \p_v \mc{K} (v,v') \partial_{v'}~. 
} 

\subsubsection{Characterization of fluctuations}
\label{sec:HCME-fluc}
The discussions till now describe only the average or typical behavior of the conserved density fields and the associated current fields. 
The equation \eqref{hcmeTev} describes the evolution of the average temperature profile as well as the evolution of a localized energy pulse in a thermally equilibrated system. 
However,  other interesting aspects that  characterize the NESS are the distributions of density and current fluctuations, long range correlations and the large deviations. To study these aspects, one requires to have a stochastic description of the evolution at the macroscopic length and time scales. 

In the context of diffusive transport, a general framework called the macroscopic fluctuation theory has been developed in the last decade which allows to provide such a  description for fluctuations \cite{Derrida07, Bertini15, Bertini02}. Starting from the microscopic description of the system one can show that in the diffusive scaling limit, the fluctuating energy density field $e(x,t)$  and the corresponding fluctuating current $J_e(x,t)$  still satisfy the continuity equation but now, in addition to the regular diffusive part of the current, there is a fluctuating part $J_e(x,t)=-D(e) \frac{\partial e(x,t)}{\partial x}+\sqrt{\chi(e)}~\eta(x,t),$
where  $\chi(e(x,t))$ is  the  mobility of the system which is related to the diffusivity $D(e(x,t))$ through the fluctuation dissipation relation and $\eta(x,t)$ is a mean zero white Gaussian noise with the properties $\langle \eta(x,t)\rangle=0$ and $\langle \eta(x,t)\eta(x',t')\rangle=\delta(x-x')\delta(t-t')$. The  evolution equation for the energy density is given by    
\begin{align}
\frac{\partial e(x,t)}{\partial t} = \frac{\partial }{\partial x} \left[D(e) \frac{\partial }{\partial x}e(x,t) -\sqrt{\chi(e)}~\eta(x,t) \right]. \label{diff-HD} 
\end{align}
Starting from this stochastic equation one can compute various moments, fluctuations  and correlations of $e(x,t)$ and $j(x,t)$ both in stationary and non-stationary regime. This description also allows one to compute the probabilities of observing atypical density and current profiles which are characterized by large deviation functions.  The whole program has been established and applied in several microscopic systems which show diffusive behavior at macroscopic scales. 
We ask if a similar procedure works for our system, displaying anomalous transport, and described by the fractional diffusion equation. Recently such an extension has been proposed in \cite{Kundu19} which we now describe.
The approach in \cite{Kundu19} is to include a noise part in the current expression in such a way that the fluctuation-dissipation theorem is satisfied. For a system in equilibrium at temperature $T$ this leads to the unique choice 
\eqa{ \begin{split}
\p_\tau e(v,\tau) &=- \p_v j(v,\tau), \\ 
{\rm with}~~ j(v,\tau) &= -\bar{\kappa} \int_0^1 dv' \mc{K}(v,v') \p_{v'}{e(v',\tau)} - \sqrt{2\bar{\kappa}} T \int_0^1 dv' B(v,v')  \eta(v',\tau)  \end{split}
\label{eq:fluccurrent}
}
where $\eta(v,\tau)$ is white Gaussian noise with $\blang \eta(v,\tau)\brang=0$, $\blang \eta(v,\tau) \eta(v',\tau') \brang = \delta(v-v')\delta(\tau-\tau')$ and the fluctuation-dissipation theorem  implies the relation
\eqa{
\mc{K}(v,v')=BB^\dagger (v,v')~,
}
with $B^\dagger$ defined as the adjoint of $B$. It was verified in \cite{Kundu19} that  Eq.~(\ref{eq:fluccurrent})  reproduces correctly  results on  energy correlations and current fluctuations in equilibrium. Extending this approach to the non-equilibrium situation was also attempted in \cite{Kundu19} and a conjecture for long-range correlations in the NESS was proposed.

\subsubsection{Role of boundary conditions: Hydrodynamic theory}
\label{sec:BC}
In the previous section we have mainly discussed the fixed boundary condition, in which case we have learned that the transport behavior in HCME model is anomalous with exponent $\alpha=1/2$ and the Fourier's law gets modified to a non-local linear response relation as in the form of Eq.~\eqref{j-nonL-hcme} with an explicit form for the kernel $\mc{K}(v,v')$ given in Eq.~\eqref{defKhcmeFi}. Also in this case the evolution of the temperature profile is given by a non-local equation \eqref{hcmeTev}  with $\mb{L}^p$ defined through Eqs.~\eqref{defKhcmeFi} and \eqref{mbL-K-rel}. In this section we would like to understand the  dependence of these results on the choice of boundary conditions. In particular we focus on the case of free boundary conditions \emph{i.e.} for $\zeta=1$ in Eq.~\eqref{eom}.

Energy transport in HCME with free boundary condition was studied numerically in \cite{Delfini10} where it was observed that the system size scaling of the current $j$ in the steady state is again proportional to $1/\sqrt{N}$, as for fixed BC. However, in contrast to the fixed BC case,  the proportionality constant depends on the friction coefficient $\lambda$.
It was also observed that the temperature profile in this case is non-linear but the associated meniscus exponent $\mu$ depends strongly on the relative values of $\lambda$ and $\omega$. For this case finding the appropriate boundary conditions for Eqs.~(\ref{PDE-hcmea},\ref{PDE-hcmeb},\ref{PDE-hcmec}) is a difficult problem \cite{Delfini10} and has so far not been possible. A different approach, based on linear response theory and NFH was proposed in \cite{Cividini17} and we present some details here.  

This approach starts with the following non-local linear response result  
\begin{align}
j(x)=-\int_0^N dy~K_N(x,y)~\frac{d T(y)}{d y},~~\label{LR-ano}
\end{align} 
which is based on a linear response calculation as done in \cite{Kundu09} but around a local equilibrium state characterized by a temperature profile. According to this calculation the Kernel is related to the equilibrium current-current correlation  \cite{Cividini17} 
\eqa{
K_N(x,y)=\frac{1}{\bar{T}^2} \int_0^{aN} dt~\langle j(x,t)j(y,0) \rangle_{eq}, \label{eq:defK}
}
where $j(x,t)$ is the  local current and $a$ is a constant. For systems with AHT we expect $N \langle j(x,t)j(y,0) \rangle_{eq} \sim t^{1-\alpha}$ which means that $K_N(x,y)$ should scale as $N^{\alpha-1}$. Hence we expect that the limit 
\eqa{
\mc{K}(v,v')=\lim_{N \to \infty} N^{1-\alpha} K_N(vN,v'N), \label{K_lim}
}
exists, which implies also that $j=J/N^{1-\alpha}$ with $J$ given by 
\eqa{
{J}=-\Delta T \int_0^1dv'~\mc{K}(v,v')~\p_{v'}\Theta(v'). \label{j_const}
}
where the temperature profile $T(x)$ is assumed to have the scaling form $T(x)=\bar{T}+\Delta T ~\Theta(x/N)$. This equation can then be used to compute the NESS temperature profile and also the current. The remaining task now is to compute the kernel $\mc{K}(v,v')$.

For HCME, the kernel $\mc{K}(u,v)$ has recently been computed in \cite{Cividini17} using the techniques of NFH as introduced in Sec.~\eqref{sec:NFH}. Following this procedure for the HCME model, one finds that  on hydrodynamic length and time scales, a random fluctuation created inside the system decomposes into two ballistically moving but diffusively spreading sound modes $\phi_\pm$ and a stationary heat mode $\phi_0$. In terms of the local stretch $r_i=q_{i+1}-q_i$ and energy $e_i=p_i^2/2 + \omega^2 r_i^2/2$, the sound modes and the heat mode are expressed as $\phi_\pm=\omega r \mp p$ and $\phi_0=e$, respectively. The evolution of these modes are given by \cite{Lepri16}
\begin{eqnarray}
\label{eq:hydro}
\partial_t \phi_\pm &=& - \partial_x [\pm c_s \phi_\pm - D \partial_x \phi_\pm - \sqrt{2 D} \eta_\pm], \\
\partial_t \phi_0 &=& - \partial_x [ G(\phi_+^2-\phi_-^2) - D_0 \partial_x \phi_0 - \sqrt{2 D_0} \eta_0], \nonumber
\end{eqnarray}
where $c_s = \omega$ is the speed of sound, $\eta_+$, $\eta_0$
and $\eta_-$ are uncorrelated Gaussian white noises, 
$G = \frac{\omega}{4}$  and $D$ and $D_0$ are phenomenological diffusion coefficients. 

The instantaneous energy current can be read from~\eqref{eq:hydro},
\begin{equation}
\label{eq:defJ}
j(x,t) = G [\phip^2(x,t)-\phim^2(x,t)],
\end{equation}
neglecting the sub-dominant terms arising from the momentum exchange and the noises  $\etapm$ ~\cite{Spohn15}. The stochastic momentum exchange process generate  a diffusive contribution [see Eq.~\eqref{eq:hydro}] which becomes sub-leading at large $N$ and the noises $\etapm$ also do not contribute since their time averages vanish.

In order to compute the  the kernel in~\eqref{eq:defK} using the form of $j(x,t)$ in \eqref{eq:defJ}, one needs to solve the equations of $\phi_\pm$ in \eqref{eq:hydro} inside a finite domain with suitable BCs. 
At this point we would like to mention that originally the NFH theory was formulated for an infinite domain \cite{Spohn15}. The work in \cite{Cividini17} provides an extension to incorporate boundary conditions for a finite domain, in the context of  the HCME model. As the equations for $\phip$ and $\phim$ are independent of $\phiz$, it is straightforward to write the solution in terms of the appropriate Green's function, as shown later.

We now discuss how to get the boundary conditions of fields $\phi_\pm$. The strategy that has been followed in \cite{Cividini17} is to introduce extra stretch and momentum variables in such a way that the equations at the boundary points ($i=1,N$) are also included into the structure of the bulk equations. This can be acheived by introducing additional conditions, which  after appropriate coarse-graining become the hydrodynamic BCs . 
To explain the procedure let us consider the free BC case as an example. 
We first introduce an extra dynamical variable $r_0$ in such a way that the form of the equation satisfied by $p_1$ becomes same as that of the bulk evolution equations with the condition
\begin{equation}
\omega^2 r_0 = \lambda p_1,
\end{equation}
where we have neglected the noise terms in \eqref{eom}. This provides one BC. We need another BC 
as the Eq.~\eqref{eq:hydro} is of second order in space. As before, introducing $p_0$ in such a way that one can make $r_0$ to satisfy a regular equation of motion in the bulk at the cost of an extra condition, provides the second BC.  Taking single derivative with respect to time on both sides of the first condition yields 
\begin{equation}
p_1 - p_0 = \lambda (r_1 - r_0).
\end{equation}
One can get two other boundary conditions by applying similar procedure to the equations of the last ($N$th) particle. Finally, coarse-graining over space and expressing the stretch $r$ and momenta $p$ in terms of the sound modes $\phipm$, we obtain  the  following BCs for free boundaries:
\begin{eqnarray}
\label{eq:bcsfree}
(\partial_x \phip + w~\partial_x \phim) |_{x=0} = ( \phip - w~\phim) |_{x=0} &=& 0, \nonumber \\
(\partial_x \phim + w~\partial_x \phip) |_{x=N} = ( \phim - w~\phip) |_{x=N} &=& 0
\end{eqnarray} 
where 
\begin{equation}
\label{eq:r}
w=\frac{\lambda-\omega}{\lambda+\omega}.
\end{equation}
These BCs can be  interpreted physically as some sort of partially `reflecting' boundaries. 
The BCs on the first (second) line of Eq.~\eqref{eq:bcsfree} mean that when a $\phip$ (resp. $\phim$) Gaussian peak hits the right (resp. left) boundary, it gets reflected as a $\phim$ (resp. $\phip$) Gaussian peak with area under the peak reduced  by a factor $w$.  This feature has been observed in numerical simulations and the validity of \eqref{eq:bcsfree} has been confirmed \cite{Cividini17}. There are two interesting cases $w=0$ and $w \to1$. In case of \textit{resonance} (also called impedance matching) $\lambda=\omega$ \emph{i.e.} $w=0$ \cite{Lepri11}, once a $\phi_\pm$ peak hits  the boundary nothing gets reflected because everything gets absorbed at the boundary reservoirs. On the other hand, $w\to1$ corresponds to almost perfectly reflecting case. This situation arises for the fixed BCs in the microscopic dynamics. Following a similar procedure as done for free BCs, it is possible to show that one arrives at the same hydrodynamic BCs Eq.~\eqref{eq:bcsfree} except now $w =1$. From Eq.~\eqref{eq:r}, one can easily see that the $w \to 1$ limit is achieved for $\lambda \to \infty$. In this limit, the $1$st and the $N$th particles hardly move \emph{i.e.} their positions $q_1$ and $q_N$ stay very close to $0$ for all times due to infinite dissipation and therefore mimic the fixed BCs for the microscopic dynamics. So for fixed BCs we  have the hydrodynamic BCs Eq.~\eqref{eq:bcsfree} with $w=1$. 

Since the hydrodynamic equations~\eqref{eq:hydro} for $\phip$ and $\phim$ along with the BCs~\eqref{eq:bcsfree}  are linear, it is easy to solve them for arbitrary initial condition. The solutions are best expressed in terms of the four Green's functions $f_{\sigma, \tau}(x,y,t)$ for $\sigma,\tau=\pm$, as 
\begin{eqnarray}
\label{eq:phipm}
&&\phi_\sigma(x,t) =  \sum_{\tau = \pm} \left[ \int_{y=0}^N \dd y f_{\sigma,\tau}(x,y,t) \phi_\tau(y,0)  + \sqrt{2 D} \int_{y=0}^N  \dd y \int_{t'=0}^t \dd t' f_{\sigma,\tau}(x,y,t-t') \partial_{y} \eta_\tau(y,t') \right], \\
&&	 \text{where,}~~~f_{\sigma,\tau}(x,y,t) = \sum_{n=-\infty}^{\infty} w^{2 n + \frac{\sigma-\tau}{2}}
 \frac{\exp \left(-\frac{(x- \sigma \tau y + 2 \sigma n N -\sigma \cs t)^2}{4 D t}\right)}{\sqrt{4 \pi D t}}, 
\label{eq:greenf}
\end{eqnarray}
with $w=1$ for fixed BCs and $w=\frac{\lambda-\omega}{\lambda+\omega}$ for free BCs.

Using this expression in Eq.~\eqref{eq:defJ} one finally gets from Eqs.~(\ref{eq:defK},\ref{K_lim}) the following expression for the kernel:
	\begin{eqnarray}
	\label{eq:solK}
\mc{K}(v,v') &=& A~{\mc{K}}_R,~~\text{where}~~\mc{K}_R=\frac{1}{\sqrt{2 \pi}}\sum_{n=-\infty}^{\infty} \left[  \frac{R^{|2 n|}}{\sqrt{|2 n+ v-v'|}} - \frac{R^{|2 n+1|}}{\sqrt{|2 n +v +v'|}} \right],
	\end{eqnarray}
where the constant $A=\frac{G^2 S^2}{ \bar{T}^2 \sqrt{ D \cs}}$ with $S = \langle \phip(x,0)^2\rangle_\mathrm{eq} = \langle \phim(x,0)^2\rangle_\mathrm{eq}=2 \bar{T}$ and $R=w^2$. The diffusion constant $D$ appearing in the equation for $\phi_\pm$ arises from the exchange mechanism and it can be shown from a microscopic calculation that $D=\gamma/2$. This then gives $A=\omega^{3/2}/(2 \sqrt{2 \gamma})$ which we note coincides with the expression for $\bar{\kappa}$ in Eq.~\eqref{frachcme}, and so we identify $A=\bar{\kappa}$. 
One can use this kernel in Eq.~\eqref{j_const} to compute the current and the temperature profile $\Theta(v)$. 

Let us define the Greens function, $\mc{G}_R$, corresponding to the kernel $\mc{K}_R$ through the equation
\eqa{\label{greenKR}
\int_0^1 dv'' \mc{K}_R(v,v'')\mc{G}_R(v'',v') = \delta(v-v').
} 
Then Eq.\eqref{j_const} can be inverted to give
\eqa{\label{greendT}
\p_v \Theta(v) = -\frac{J}{\bar{\kappa}\Delta T} \int_0^1 dv' \mc{G}_R(v,v')
}
Solving this equation with the boundary conditions $\Theta(0)=1/2,\Theta(1)=-1/2$ gives us the expressions for the current and temperature profile
\eqa{\label{greenJ}
J&=\bar{\kappa} \Delta T \left[\int_0^1 \int_0^1 dv' dv'' \mc{G}_R(v',v'')\right]^{-1}~, \\
\Theta(v)&=\frac{1}{2}-\frac{ \int_0^v \int_0^1 dv' dv'' \mc{G}_R(v',v'')}{\int_0^1 \int_0^1 dv' dv'' \mc{G}_R(v',v'')}~. \label{greenT}
}
One uses this in Eq.~\eqref{j_const} to solve for the temperature profile $\Theta(v)$.  
The above analysis, based on linear response calculation, assumes $|\Delta T| << \Tb$. However for HCME, one observes that the quadratic correlations satisfy a closed set of linear equations with a source term proportional to $\Delta T$~\cite{Lepri09}. Hence the temperature profile $\Theta(v)$ in \eqref{greenT} is also valid for any $\Delta T$.

\begin{figure}[t]
	\begin{center}
		\includegraphics[width=1.0\textwidth]{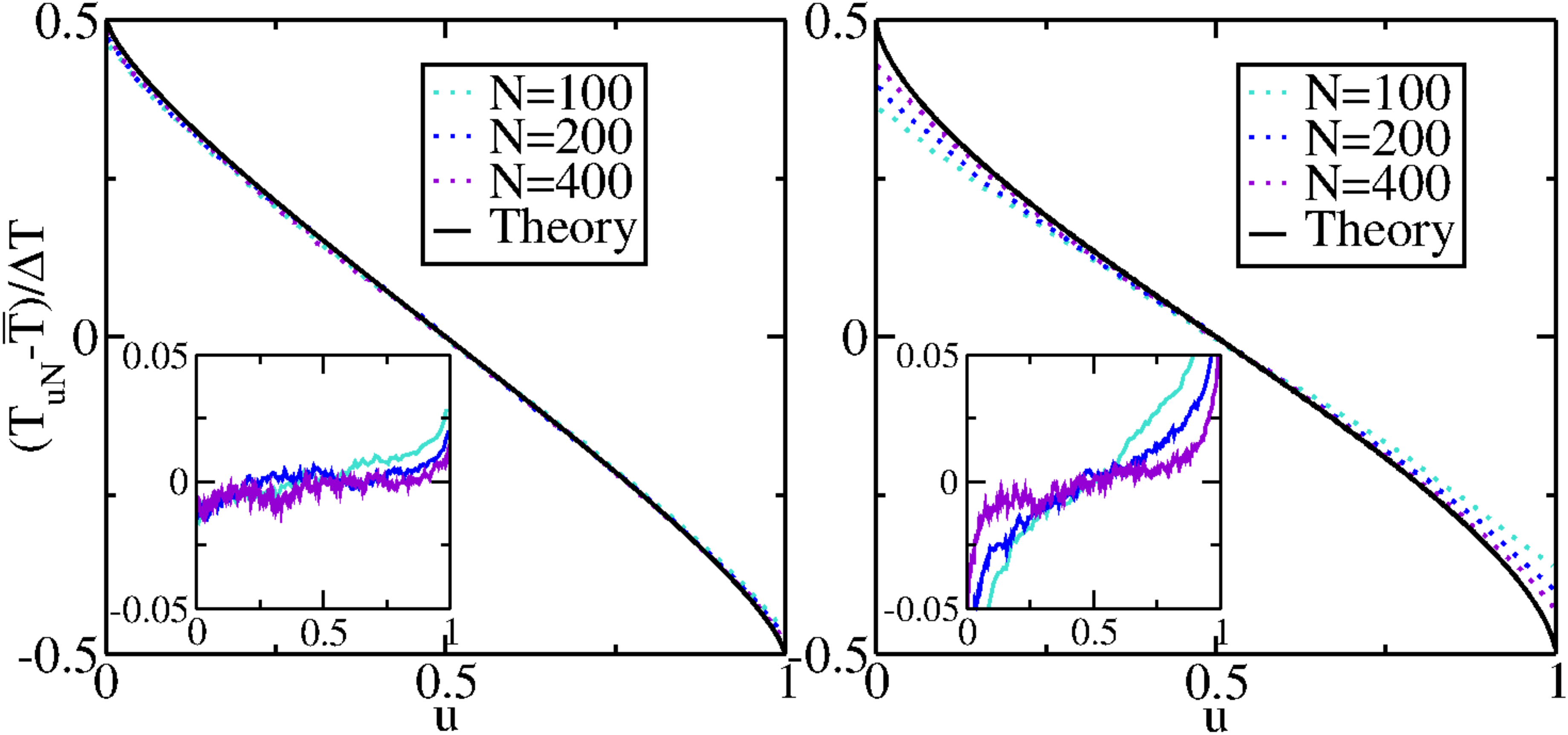} 
	\end{center}
	\caption{\small Rescaled temperature profile for resonant BCs $R=0$ (left panel) and free BCs with $R=\frac{1}{2}$ (right panel). In the main plots results of Monte-Carlo simulations for increasing system sizes $N=100$, $200$ and $400$ are compared to the theoretical predictions given by Eq.~\eqref{theta_R_0} for the $R=0$ case (left panel) and the numerical solution of
		Eq.~(\ref{j_const},\ref{eq:solK})
	 with $R=\frac{1}{2}$ for the plots in the right panel. In the insets the differences between measurements and theory are shown. The other parameter values are $\Tp=1.5$, $\Tm=0.5$, and $\omega=\gamma=1$. 
(Adapted from  Cividini et.al. with permission from \cite{Cividini17} Copyright SISSA Medialab Srl, IOP Publishing.)
}
	\label{fig:tprof}
\end{figure}

It turns out the equation \eqref{greendT} can be solved analytically and  exact expressions of the temperature profile $\Theta(v)$ can be obtained in the following two limiting cases ---

(i) Free resonant case $R=0$:  In this case the kernel is simply given $\mc{K}_0=1/\sqrt{2\pi~|v-v'|}$  which is same as that of an infinite system. For this kernel, the solution of Eq.~\eqref{j_const} can be directly written using standard results on solution of integral equations \cite{Polyanin08} as
\eqa{
\p_v \Theta(v) = -\frac{J}{\bar{\kappa}\Delta T} \frac{1}{\sqrt{\pi} v^{1/4}(1-v)^{1/4}}~.
}
This can be solved with the boundary conditions to give the temperature profile
\eqa{
\Theta_{R=0}(v)=\frac{1}{2}-\frac{\sqrt{\pi}v^{3/4}  {}_2F_1\left(\frac{1}{4},\frac{3}{4};\frac{7}{4};v\right)}{2\Gamma\left(\frac{3}{4}\right)\Gamma\left( \frac{7}{4}\right)},\label{theta_R_0}
}
where $ _2F_1$ is hypergeometric function, and the current
\eqa{
J={\bar{\kappa} \Delta T}\frac{\pi}{2 \Gamma^2(3/4)}
}
 This profile is verified numerically in Fig.~\ref{fig:tprof} (left panel), where we observe diverging derivatives at the boundaries. From the above expression it is possible to show that the meniscus exponent is $\mu=3/4$.

(ii) Perfectly reflecting case $R \to 1$: As mentioned above this is equivalent to fixed BC  for which the temperature profile, given in Eq.~\eqref{THCME-SS}, was computed from microscopic calculation in the previous section. In this case  it is known \cite{Kundu19} that the  eigenfunctions of the operator $\mc{K}_R$ are precisely the sine-functions $\alpha_n(v)$, \emph{i.e} 
\eqa{
\int_0^1 dv' \mc{K}_R(v,v') \alpha_n(v') = (n \pi)^{-1/2} \alpha_n(v)~,
}
which is consistent with Eq.~\eqref{defKhcmeFi}. This then gives us the corresponding Green's function
\eqa{
\mc{G}_R(v,v')= \sum_{n=1}^\infty (n \pi)^{1/2}\alpha_n(v) \alpha_n(v')~. 
}
Using this and Eqs.~(\ref{greendT},\ref{greenJ}) we recover the exact expressions for the temperature profile and current given in Eqs.~(\ref{THCME-SS},\ref{JHCME-SS}) \cite{Kundu19}.

For free BCs with $\lambda \neq \omega$ we have $0 < R < 1$. In this case it is difficult to solve Eqs.~(\ref{j_const},\ref{eq:solK}) analytically but  numerical solutions have been obtained. In Fig.~\ref{fig:tprof} (right panel) a comparison of the temperature profile obtained from the numerical solution and from direct microscopic simulations for $R=1/2$ and one can observe excellent agreement. Note again that the temperature profile is singular at the boundaries. It turns out that the exponent $\mu$ characterizing this singularity depends on not only on $\alpha$ but also on $R$~\cite{Lepri11}.
\begin{figure}[t]
	\begin{center}
		\includegraphics[width=0.75\textwidth]{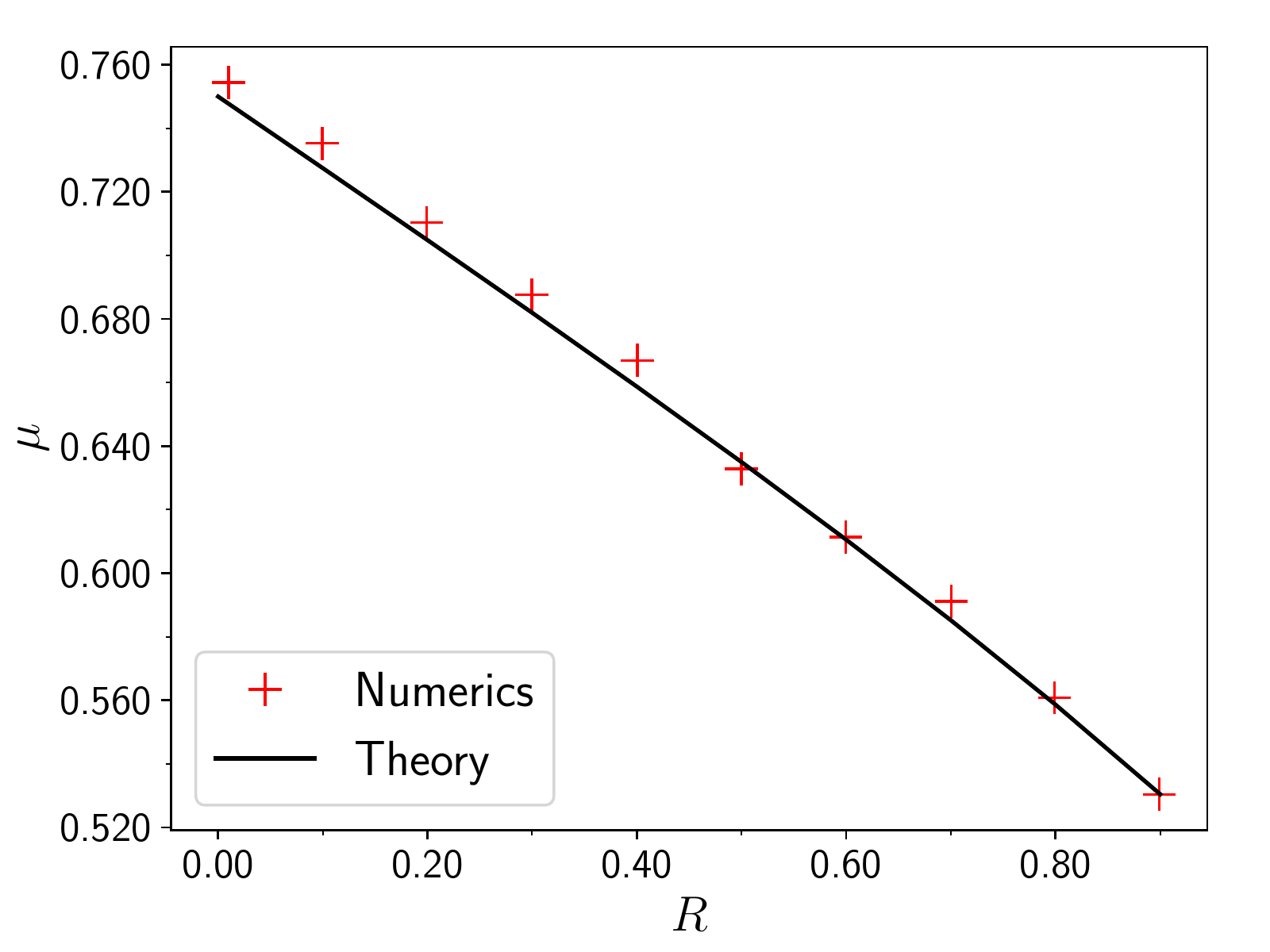} 
	\end{center}
	\caption{\small The meniscus exponent $\mu$ as a function of $R$ for $\alpha = 1/2$. The prediction from~\eqref{eq:nuimp2} is plotted against the values obtained from the numerical solution of the integral equation~\eqref{j_const} with the kernel in Eq.~\eqref{eq:solK}.}
	\label{fig:Rnu}
\end{figure}
To determine this dependence we take a derivative with respect to $v$  of Eq.~\eqref{j_const} and get $\int_0^1dv'~\partial_v\mc{K}_R~\partial_{v'}\Theta(v')=0$. Although the integral is identically zero for all $v$, the individual terms in the integrand have divergences. For example, the kernel diverges as $\mc{K}_R\sim |v-v'|^{-1/2}$ while $\partial_{v'}\Theta(v')$ diverges as $|\delta v|^{\mu-1}$.  Requiring that all divergent integrals cancel each other, leads to the following relation between $R$ and $\mu$:
\begin{equation}
\label{eq:nuimpl}
R=\frac{\int_{0}^1 \frac{q^{\mu-1}-q^{1/2-\mu}}{(1-q)^{3/2}} \dd q}{\int_{0}^1 \frac{q^{\mu-1}+q^{1/2-\mu}}{(1+q)^{3/2}} \dd q}~. 
\end{equation}
The integrals can be performed explicitly to give 
\eqa{
\label{eq:nuimp2}
\mu=1-\frac{1}{\pi}\rm{arctan}\left(\frac{\sqrt{2-R^2}+R}{\sqrt{2-R^2}-R}\right), 
}
which is plotted in Fig.~\ref{fig:Rnu} along with  results extracted from the temperature profile  obtained from direct numerical  solution of Eq.~\eqref{j_const}. We note that this result differs from the one conjectured in \cite{Lepri11} though rather interestingly, the values of $\mu$ at $R=0$ and $R=1$ obtained from the two expressions agree. A generalization of the above result for arbitrary $\alpha$ is possible using the  L\'{e}vy walk approach with the general kernel
 in Eq.~\eqref{LWchiR} which leads to an expression similar to Eq.~\eqref{eq:nuimpl}, now with the right hand depending explicitly on $\alpha$.

\section{Summary and open questions}
\label{sec:summary}
Heat transport in a large class of one-dimensional systems with Hamiltonian or conservative stochastic dynamics is known to be anomalous. Some typical  signatures of anomalous transport include NESS studies which find that the thermal conductivity $\kappa$, diverges with system  size $N$ as  $\kappa \sim N^\alpha$, and the temperature profile $T(x)$ is typically nonlinear, with a singular dependence $T(\delta x) \sim (\delta x)^\mu$ for small distance $\delta x$ from the boundary. In the closed system one finds that heat pulses and correlation functions spread super-diffusively and are associated to propagators  that have the scaling form $t^{-\gamma} \mc{G}(t^{-\gamma}x)$. The scaling form is valid for times $|x| < ct$, where $c$ is the sound speed in the system, beyond which time the correlations decay exponentially. The scaling function is given by the  L\'{e}vy-stable distribution in the bulk and the finite cut-off leads to the width of the pulse scaling   as $\sigma(t) \sim t^{\beta/2}$. 

In this review we discussed these signatures of anomalous transport and showed how they can be understood within three different but related frameworks --- (a) a phenomenological model where the heat carriers are taken to be  L\'{e}vy walkers, (b) a microscopic phenomenological approach based on nonlinear fluctuating hydrodynamics and (c) exact results obtained for certain stochastic models. The main picture that emerges is that anomalous heat transport can be understood by replacing Fourier's law in Eq.~\eqref{FL} by a non-local fractional-type diffusion equation given in Eq.~\eqref{NLFL}, where the precise form of the kernel $K_R(x,y)$ depends on the specific set-up and boundary conditions. For the stochastic models the form of the kernel is known explicitly both for the closed system (infinite line) and the open system. In the  L\'{e}vy walk picture, where the distribution of flight times has a power-law dependence $\sim 1/t^{\nu +1}$, the kernel has the asymptotic form $K_R(x,y) \sim 1/|x-y|^{\nu-1}$. We saw from the various approaches, that all the different exponents mentioned above are related to each other and in fact can be expressed in terms of the  L\'{e}vy walk exponent as
\eqa{
\alpha=2-\nu,~~~\beta=3-\nu,~~~ \gamma=1/\nu~. 
}
For the Hamiltonian models that we discussed, namely the alternate mass hard-particle gas and the FPUT model, the various exponents are given by $\alpha=1/3, \beta=4/3, \gamma=3/5$ and correspond to a  L\'{e}vy-walk exponent $\nu=5/3$. For the stochastic momentum exchange model we have  $\alpha=1/2, \beta=3/2, \gamma=2/3$ which corresponds to $\nu=3/2$. 
The meniscus exponent $\mu$ is non-universal and depends on $\nu$ and on boundary conditions through a single dimensionless number $R$, which can be interpreted as the reflection coefficient of the  L\'{e}vy walkers at the boundaries. 
In the context of the exactly solvable stochastic models, we discussed the spectrum of the fractional-type Laplacian operator [specified by the kernel $K_R(x,y)$] in the open set-up, and pointed out important differences with the spectrum of the usual Laplacian for diffusive processes.

We conclude by mentioning some outstanding open questions in the field. 

\begin{itemize}
\item  Hamiltonian systems ---  The  L\'{e}vy walk behaviour has been clearly observed in large number of simulations. The formalism of NFH gives a microscopic justification of the  L\'{e}vy walk model and the fractional-diffusion type description of the heat mode. Some open questions include:
\begin{enumerate}
\item A more rigorous microscopic derivation of the evolution equation of a localized heat pulse in  an equilibrium system, to show that the central peak satisfies a fractional-diffusion type equation of a  form  similar to that in Eq.~\eqref{NLFL}.
\item Extension of the NFH formalism to the non-equilibrium case to study transport in  finite open system and understand the role of BCs. Detailed simulations are also required to understand the effect of BCs.
\item Establishing the  L\'{e}vy walk picture from a microscopic viewpoint ?
\end{enumerate}
\item  Stochastic systems ---
For the HCME model, the non-local version of Fourier's law has been established and the response kernel $K_R$ computed so far using two methods:  (i) exact microscopic method for the BC corresponding to $R=1$ and (ii) using NFH for arbitrary $R$. Is it possible to extend the exact microscopic approach to find the non-local kernel $K_R$ for general boundary conditions. Similarly for the HCVE it would be interesting to explore the role of BCs.
\item For the HCME model, it has been possible to find the eigenspectrum of the non-local kernel $K_R$ for the case $R=1$ and it was observed that the eigenvalues for Dirichlet and Neumann boundary conditions differ (unlike for the usual Laplacian). Finding the spectrum of the non-local kernel $K_R$ for general $R$, for Dirichlet and Neumann boundary conditions, is an interesting mathematical problem. The knowledge of the spectrum, namely eigenvectors and eigenvalues, enables one to study the time-evolution.   
\item For the HCME model we showed that it is possible to write a stochastic non-local equation [Eq.~\eqref{eq:fluccurrent}] to describe equilibrium fluctuations. An open problem  is to write such an equation in the  non-equilibrium set-up. For diffusive systems this is given by Eq.~\eqref{diff-HD} and this equation enables  one to compute long-range correlations in the NESS and large deviation functions.
\end{itemize}

\vskip 1cm

{\bf Acknowledgments}:  Anupam Kundu  would like to acknowledge the support from DST grant under project No. ECR/2017/000634. Abhishek Dhar would like to acknowldege support of the project EDNHS ANR-14-CE25-0011 of the French National Research Agency (ANR). Anupam Kundu and Abhishek Dhar  would like to acknowledge the support from the project 5604-2 of the Indo-French Centre for the Promotion of Advanced Research (IFCPAR).


\end{document}